\documentclass[aps,preprint,superscriptaddress,nofootinbib,preprintnumbers,eqsecnum,prd]{revtex4-1}

\usepackage{bm,hyperref,amsmath,amssymb,amsfonts,slashed,array,graphicx,soul,multirow,mathtools,textcomp}
\usepackage[vcentermath,noautoscale]{youngtab}

\Yboxdim{8pt}

\newcommand{\Tr}{\text{Tr}}

\newcommand{\G}{\Gamma}
\newcommand{\g}{\gamma}
\newcommand{\Br}{\text{Br}}
\newcommand{\LambdaP}{\Lambda_{\text{portal}}}

\newcommand{\tPi}{{\hat{\Pi}}}
\newcommand{\tzero}{{\hat{\pi}^0}}
\newcommand{\tplus}{{\hat{\pi}^+}}
\newcommand{\tminus}{{\hat{\pi}^-}}
\newcommand{\tpm}{{\hat{\pi}^\pm}}
\newcommand{\tmp}{{\hat{\pi}^\mp}}
\newcommand{\szero}{{\hat{\eta}}}
\newcommand{\hpi}{{\hat{\pi}_2}}
\newcommand{\lpi}{{\hat{\pi}_1}}
\newcommand{\lhpi}{{\hat{\pi}_{1,2}}}

\newcommand{\hc}{\text{ h.c.}}

\newcommand{\Sref}[1]{Sec.~\ref{#1}}

\def\GeV{\text{GeV}}
\def\TeV{\text{TeV}}

\renewcommand{\tilde}[1]{\smash{\overset{\mbox{\texttildelow}}{\smash{#1}\phantom{\rule{2pt}{1.4ex}}}}\hspace{-2pt}}

\makeatletter
\g@addto@macro\bfseries{\boldmath}
\makeatother

\allowdisplaybreaks
\usepackage[usenames,dvipsnames,table]{xcolor}

\usepackage{array}
\newcolumntype{C}[1]{>{\centering\let\newline\\\arraybackslash\hspace{0pt}}m{#1}}

\begin{document}
\preprint{FERMILAB-PUB-16-089-T}
\title{Diphotons from an Electroweak Triplet-Singlet}
\author{Kiel Howe}
\affiliation{Theoretical Physics Department, Fermi National Accelerator Laboratory, Batavia, IL 60510 USA}
\author{Simon Knapen}
\author{Dean J. Robinson}
\affiliation{Department of Physics, University of California, Berkeley, CA 94720, USA}
\affiliation{Ernest Orlando Lawrence Berkeley National Laboratory,
University of California, Berkeley, CA 94720, USA}

\begin{abstract}
The neutral component of a real pseudoscalar electroweak (EW) triplet can produce a diphoton excess at 750\,GeV, if it is somewhat mixed with an EW singlet pseudoscalar. This triplet-singlet mixing allows for greater freedom in the diboson branching ratios than the singlet-only case, but it is still possible to probe the parameter space extensively with 300\,fb$^{-1}$. The charged component of the triplet is pair-produced at the LHC, which results in a striking signal in the form of a pair of $W\g$ resonances with an irreducible rate of $0.27$\,fb. Other signatures include multiboson final states from cascade decays of the triplet-singlet neutral states. A large class of composite models feature both EW singlet and triplet pseudo-Nambu Goldstone bosons in their spectrum, with the diboson couplings generated by axial anomalies.
\end{abstract}

\maketitle
\clearpage

\tableofcontents
\clearpage

\section{Introduction}
Decays of exotic states to standard model (SM) vector bosons may produce striking signatures at the LHC. A hint of a diphoton resonance with mass nearby 750\,GeV and rate $\sim 5$\,fb \cite{ATLAS-CONF-2015-081,CMS-PAS-EXO-15-004} has prompted an extensive bombardment of the Literature, containing copious investigations of both the phenomenology and possible sources of such a signal. Embedding this signature into a consistent theory leads to expectations for signatures in other decay channels, in particular the diboson channels $\gamma\gamma$, $ZZ$, $Z\gamma$, and $W^+ W^-$, as well as various exotic decay channels and associated production modes.

If the source of this signal is a (pseudo)scalar, the  simplest scenario  is an electroweak (EW) singlet~\cite{Harigaya:2015ezk,Mambrini:2015wyu,Backovic:2015fnp,Angelescu:2015uiz,Nakai:2015ptz,Knapen:2015dap,Buttazzo:2015txu,Pilaftsis:2015ycr,Franceschini:2015kwy,McDermott:2015sck,Ellis:2015oso,Low:2015qep,Bellazzini:2015nxw,Gupta:2015zzs,Molinaro:2015cwg,Bian:2015kjt,Agrawal:2015dbf,Falkowski:2015swt,Aloni:2015mxa,Bai:2015nbs,Altmannshofer:2015xfo}. In the presence of CP conservation, such an EW pseudoscalar singlet, $\szero$, may decay to diboson final states via the usual dimension-five field strength operators, $\szero B_{\mu\nu} \tilde{B}^{\mu\nu}$ and $\szero\Tr[W_{\mu\nu} \tilde{W}^{\mu\nu}]$, without requiring additional sources of electroweak symmetry breaking (EWSB), and without mixing with the SM Higgs. Moreover, it may be produced abundantly by gluon fusion, via $\szero G_{\mu\nu} \tilde{G}^{\mu\nu}$. The presence of such a pseudoscalar in Nature therefore can account for the diphoton excess, while remaining consistent with Higgs coupling measurements and electroweak precision observables.

In this work, we extend this scenario to include the next lowest $SU(2)_L$ electroweak representation with these properties: A pseudoscalar triplet $\tPi \sim(\tzero,\tpm)$ furnishing the $\bm{3}_0$ of $SU(2)_L \times U(1)_Y$. The components of this triplet may decay to diboson final states via the dimension-5 field strength operator $B_{\mu\nu} \Tr\{\tPi \tilde{W}^{\mu\nu}\}$ (the other dimension-5 operator $\Tr\{\tPi W_{\mu\nu} \tilde{W}^{\mu\nu}\}$ is identically zero). Such an EW triplet need not acquire an EWSB vacuum expectation value in order to decay and its neutral component does not mix with the SM Higgs if CP is conserved. Since the SM Higgs remains the only source of spontaneous EWSB, this scenario is intrinsically different from models where the neutral component of an $SU(2)_L$ doublet is responsible for the diphoton excess~\cite{Angelescu:2015uiz,Gupta:2015zzs,Becirevic:2015fmu,Falkowski:2015swt,Aloni:2015mxa,Altmannshofer:2015xfo}, as well as from left-right symmetric approaches to the diphoton excess~\cite{Dey:2015bur,Berlin:2016hqw,Borah:2016uoi,Dasgupta:2015pbr,Staub:2016dxq,Ren:2016gyg} or Georgi-Machacek models~\cite{Fabbrichesi:2016alj,Chiang:2016ydx}, in which an EW triplet acquires a vev, and from extensions of $SU(2)_L$~\cite{Boucenna:2015pav,Hernandez:2015ywg,Cao:2015scs}. 

The SM Higgs EWSB vev, $v$, induces $\tzero$--$\szero$ mixing at $\mathcal{O}(v^2)$, opening up a sizable gluon fusion production channel for both neutral mass eigenstates in the triplet-singlet admixture.\footnote{Single production of a pure EW triplet requires either photon fusion~\cite{Fichet:2015vvy,Csaki:2015vek,Csaki:2016raa,Harland-Lang:2016qjy} or vector boson fusion. These production channels are typically barely sufficient to produce the observed diphoton resonance rate without a large 't-Hooft coupling, that in turn requires the presence of a large number of flavors of exotic hypercharged states.} Compared to the pure singlet case, the triplet-singlet framework has two novel features. First, this framework admits more flexible diboson branching ratio relations. We show these relations may nevertheless be conveniently parametrized on a compact two-dimensional space together with the current and projected LHC reach. Much of this parameter space can be probed with $300$\,fb$^{-1}$. Second, pair production of charged triplet states, $q\bar{q} \to \tpm\tmp$ or $\pi^\pm\pi^0$, has a minimum rate from Drell-Yan processes that is fixed by SM EW couplings, and produces striking 4-boson signals. The phenomenology of pair production of a pure EW triplet decaying to dibosons at the LHC has been explored in Refs.~\cite{Kilic:2010et,Freitas:2010ht} with a focus on the $(W\gamma)(\gamma\gamma)$ channel. In the triplet-singlet framework this channel can be diluted by dijet decays of the neutral state, but the promising $q\bar{q} \to \tpm\tmp\to (W\gamma)(W\gamma)$ channel has an irreducible rate of  $0.27$\,fb.

We show in this paper that this EW triplet-singlet mixing scenario has a broad region of parameter space consistent with the claimed diphoton excess. It is viable if the two neutral mass eigenstates have a small mass splitting, such that they produce unresolved overlapping resonances that mimic a much broader resonance, or if they at least feature a mass splitting smaller than the $W$ mass. Mass splittings larger than the $W$ or Higgs mass open up an alternate possibility for diphoton resonance production from tree level cascade decays. However, this scenario is now in some tension with observed $p_T$ distributions and ($b$-)jet counts \cite{moriond}.

A well-motivated class of theories that can exhibit a triplet-singlet spectrum of states are vector-like composite theories, in which the EW triplet and singlet are light pseudo-Nambu Goldstone bosons (pNGBs)~\cite{Kilic:2008pm,Kilic:2009mi,Bai:2010mn}, the hyper-pions of the new composite sector.  These hyper-pions generically couple to SM gauge bosons through chiral anomalies. Such theories have been recently explored in detail in the context of a pure singlet pNGB producing the $750$\,GeV diphoton resonance~\cite{Harigaya:2015ezk,Franceschini:2015kwy,Molinaro:2015cwg,Matsuzaki:2015che,Bian:2015kjt,Bai:2015nbs,Harigaya:2016pnu,Redi:2016kip,Harigaya:2016eol}. We extend a benchmark model of this kind to include a Higgs portal coupling to the SM, which generically leads to the triplet-singlet effective theory. In addition, we describe models where the triplet-singlet effective theory is obtained with the Higgs itself part of the composite sector. Such models have also have been recently studied to explain the $750$\,GeV diphoton hints, but with a pure singlet state~\cite{Belyaev:2015hgo,Low:2015qep,Bellazzini:2015nxw,No:2015bsn}.

This paper is organized as follows. In Section~\ref{sec:framework} we describe the singlet-triplet effective theory and discuss its  generic constraints and signatures.  Section~\ref{sec:diphoton} provides details on the phenomenology related to the diphoton excess for some benchmark models, followed by a detailed exploration of diboson branching ratio relations in Section~\ref{sec:BRR}.  In Section~\ref{sec:composite} we describe possible composite pNGB UV completions. 

\section{Framework and generic signatures}
\label{sec:framework}
\subsection{Gauge interactions}
Retaining terms up to dimension-5, we consider a triplet-singlet model with gauge interactions of the form
\begin{equation}
	\label{eq:gaugebasis}
	\frac{\alpha}{8 \pi} \Bigg\{ \frac{\sqrt{2}c_\tPi}{s_Wc_W f} B_{\mu\nu}  \Tr[\tPi \tilde{W}^{\mu\nu}] + \frac{c_1}{c_W^2 f} \szero B_{\mu\nu}\tilde{B}^{\mu\nu} + \frac{2c_2}{s_W^2 f} \szero \Tr[W_{\mu\nu}\tilde{W}^{\mu\nu}]\Bigg\} +  \frac{c_3}{f} \frac{\alpha_s}{8\pi}\szero G^a_{\mu\nu}\tilde{G}_a^{\mu\nu}\,.
\end{equation}
Here the dual field strength $\tilde{X}^{\mu\nu} \equiv \epsilon^{\mu\nu\rho\sigma}X_{\rho\sigma}$, $f$ is the effective field theory scale, and $c_W \equiv \cos(\theta_W)$ and $s_W \equiv \sin(\theta_W)$ denote the cosine and sine of the Weinberg angle. The pseudoscalar and vector boson triplets are canonically normalized such that
\begin{equation}
	\tPi = \begin{pmatrix} \tzero/\sqrt{2} & \tplus \\ \tminus & -\tzero/\sqrt{2} \end{pmatrix}\,,\qquad W = \frac{1}{2}\begin{pmatrix} W^3 & \sqrt{2} W^+ \\ \sqrt{2} W^- & -W^3 \end{pmatrix}\,.
\end{equation}
The triplet mass term is $(m^2_\tPi/2)\Tr\{\tPi \tPi\}$ with these conventions, and the couplings $c_{\tPi,1,2,3}$ are normalized such that gauge couplings and anticipated loop factors are factored out. Without loss of generality, we take $c_{\tPi}\geq0$ as our sign convention. We assume the triplet-singlet sector is parity conserving, so that all couplings are real. 

After EWSB, these gauge interactions become
\begin{align}
	&\frac{\alpha}{8\pi} \frac{\tzero}{f}\bigg\{a_{\gamma\gamma}F\tilde{F} + a_{ZZ} Z \tilde{Z} + a_{Z\gamma} F \tilde{Z}\bigg\}+ \bigg[\frac{\alpha}{8\pi}\frac{\tplus}{f}\bigg\{a_{W\gamma} F\tilde{W}^- + a_{WZ}Z\tilde{W}^-\bigg\} + \hc\bigg]\notag\\
	 & \qquad \qquad \qquad + \frac{\alpha}{8\pi}\frac{\szero}{f}\bigg\{b_{\gamma\gamma}F \tilde{F} + b_{ZZ} Z \tilde{Z} + b_{Z\gamma} F\tilde{Z} + b_{WW} W^-\tilde{W}^+\bigg\}\,, \label{eqn:GBC}
\end{align}
in which 
\begin{equation}
	\label{eqn:TGI}
	a_{\gamma\gamma} = - a_{ZZ} = c_{\tPi}\,,\qquad a_{Z\gamma} = 2c_{\tPi}\cot(2\theta_W)\,,\qquad a_{W\gamma} = c_{\tPi}/s_W\,,\qquad a_{WZ} = -c_{\tPi}/c_W\,,
\end{equation}
and as usual
\begin{gather}
	b_{\gamma\gamma} = c_1 + c_2\,,\qquad b_{ZZ} = c_1\tan^2(\theta_W) + c_2 \cot^2(\theta_W)\,, \notag\\ 
	b_{Z\gamma} = 2c_2 \cot(\theta_W)- 2 c_1\tan(\theta_W)\,, \qquad b_{WW} = 2c_2/s_W^2\,. \label{eqn:BC}
\end{gather}
For the triplet components, one finds the following corresponding partial widths
\begin{equation}
\label{eqn:PW}
\begin{split}
	\Gamma^\tzero_{\gamma\gamma} & = \frac{a_{\gamma\gamma}^2}{\pi f^2}\frac{\alpha^2}{64\pi^2}m_{\tzero}^3\,,\\
	\Gamma^\tzero_{Z\gamma} & = \frac{a_{Z\gamma}^2}{2\pi f^2} \frac{\alpha^2}{64\pi^2}m_{\tzero}^3\Big(1 - m_Z^2/m_\tzero^2\Big)^{3}\,,\\
	\Gamma^\tzero_{ZZ} & = \frac{a_{ZZ}^2}{\pi f^2}\frac{\alpha^2}{64\pi^2} m_{\tzero}^3\Big(1 - 4m_Z^2/m_\tzero^2\Big)^{3/2}\,,\\
	\Gamma^\tpm_{W\gamma} & = \frac{a_{W\gamma}^2}{2\pi f^2}\frac{\alpha^2}{64\pi^2}m_{\tpm}^3\Big(1 - m_W^2/m_\tpm^2\Big)^{3}\,,\\
	\Gamma^\tpm_{WZ} &  = \frac{a_{WZ}^2}{2\pi f^2}\frac{\alpha^2}{64\pi^2}m_{\tpm}^3\bigg[\bigg(1 - \frac{(m_W+m_Z)^2}{m_\tpm^2}\bigg)\bigg(1 - \frac{(m_W-m_Z)^2}{m_\tpm^2}\bigg)\bigg]^{3/2}\,,
\end{split}
\end{equation}
and similarly for the singlet decay rates, $\Gamma^\szero_{XY}$, with the replacements $a_i \to b_i$ as appropriate. The decay rates $\szero \to W^+W^-$ and  $\szero \to gg$  are further 
\begin{equation}
\label{eqn:SPW}
\begin{split}
	\Gamma^\szero_{WW} & = \frac{b_{WW}^2}{2\pi f^2}\frac{\alpha^2}{64\pi^2} m_{\szero}^3\Big(1 - 4m_W^2/m_\szero^2\Big)^{3/2}\,,\\
	\Gamma^\szero_{gg} & = 8 \frac{c_3^2}{\pi f^2}\frac{\alpha_s^2}{64\pi^2} m_{\szero}^3\,,
\end{split}
\end{equation}
respectively. Neglecting the generally small phase space corrections in eqs~\eqref{eqn:PW}, one sees from eq.~\eqref{eqn:TGI} that the relative branching fractions to diboson final states for the triplet alone are fixed fully by just the Weinberg angle,
\begin{equation}
	\Gamma^\tzero_{Z\g} / \Gamma^\tzero_{\g\g}  \simeq 2\cot^2\theta_W \simeq 0.82\,,\qquad
	\Gamma^\tzero_{ZZ} / \Gamma^\tzero_{\g\g}  \simeq 1\,,\qquad
	\Gamma^\tzero_{WW} / \Gamma^\tzero_{\g\g} \simeq 0\,. \label{eqn:PTBR}
\end{equation}

\subsection{Triplet-singlet mixing}
Since the triplet and singlet are pseudoscalars, and we insist on parity and CP conservation in the Higgs sector, there are no cubic $H^\dagger \Pi H$ nor $H^\dagger H \szero$ operators, and hence no mixings with the Higgs.  Consequently, couplings of single pseudoscalars to the SM fermions are not induced by Higgs portal interactions, and are therefore suppressed, being negligibly generated only by higher-order interactions from the UV completion (see Sec.~\ref{sec:composite}). There are, however, Higgs portal quartic terms
\begin{equation}
	\label{eq:quartics}
	\lambda H^\dagger \tPi H \szero  + \lambda_\tPi H^\dagger \tPi\tPi H + \lambda_\szero H^\dagger H \szero^2\,.
\end{equation}
The latter two terms produce small masses for the triplet and singlet, that may be neglected compared to the larger $\tPi$ and $\szero$ mass terms. Moreover, they do not break custodial symmetry
and hence do not split the $\tzero$ and $\tpm$ masses. 

The first term, however, induces a triplet-singlet mixing and consequently mass splittings too. In detail, the mass terms are
\begin{equation}
	\label{eqn:MTE}
	\frac{1}{2}\begin{pmatrix}\tzero\\\szero\end{pmatrix}^T \begin{pmatrix}m^2_\tPi&-\frac{1}{2\sqrt{2}}\lambda v^2\\-\frac{1}{2\sqrt{2}}\lambda v^2&m^2_\szero\end{pmatrix}\begin{pmatrix}\tzero\\\szero\end{pmatrix} +m^2_\tPi \tplus\tminus.
\end{equation}
Let us define
\begin{equation}
	\delta m^2 \equiv m_\szero^2 - m_\tPi^2\,, \qquad \varepsilon \equiv \lambda v^2/\sqrt{2} \,,\qquad \Delta \equiv \sqrt{(\delta m^2 )^2 + \varepsilon^2}\,,
\end{equation}
where the Higgs vev is $v/\sqrt{2}$, $v = 246$\,GeV. Writing the lighter and heavier mass eigenstates as $\lpi$ and $\hpi$ respectively, one finds mass spectrum
\begin{equation}
	m^2_{\lpi,\hpi} = \frac{1}{2}\big(m^2_{\tPi} + m^2_{\szero} \pm \Delta \big) \simeq m^2_{\szero,\tPi} \pm \varepsilon^2/(2\,\delta m^2)
\end{equation}
in the limit that $\varepsilon \ll \delta m^2$, and mixing
\begin{equation}
	\label{eqn:MSR}
	\begin{pmatrix} \lpi \\ \hpi \end{pmatrix} = \begin{pmatrix} \cos\varphi & \sin\varphi \\ -\sin\varphi & \cos\varphi \end{pmatrix} \begin{pmatrix} \tzero \\ \szero \end{pmatrix} \,,
\end{equation}
in which 
\begin{align}
	\cos \varphi & \equiv \frac{\delta m^2 + \Delta}{\sqrt{\varepsilon^2 + (\delta m^2 + \Delta)^2}} \simeq 1 - \frac{\varepsilon^2}{8(m_\hpi^2 - m_\lpi^2)^2}\,,\notag\\
	 \sin\varphi & \equiv \frac{\varepsilon}{\sqrt{\varepsilon^2 + (\delta m^2 + \Delta)^2}} \simeq \frac{\varepsilon}{2(m_\hpi^2 - m_\lpi^2)}\,, \label{eqn:MSRD}
\end{align}
again in the limit that $\varepsilon \ll \delta m^2$. Applying the rotation in eqs~\eqref{eqn:MSR} to the gauge basis couplings~\eqref{eqn:GBC}, one can immediately read off the mass eigenstate couplings to the various diboson states, and hence the consequent partial widths from eqs~\eqref{eqn:PW}. For instance, the $\lpi \to \g\g$ partial width is
\begin{equation}\label{eq:interference}
	\Gamma^{\lpi}_{\g\g} = \big[a_{\g\g}\cos\varphi  + b_{\g\g}\sin \varphi\big]^2\frac{\alpha^2}{64\pi^3 f^2}m_{\lpi}^3\,.
\end{equation}

Hereafter, we parametrize the triplet-singlet theory in terms of the physical parameters $m_{\lpi}$, $m_{\hpi}$ and $\sin\varphi$. Note that in terms of these parameters, the underlying parameters
\begin{gather}
	m_{\tpm} = m_{\tPi}^2 = m_{\lpi}^2 \cos^2\varphi + m_{\hpi}^2 \sin^2\varphi\,,\qquad m_{\szero}^2 = m_{\lpi}^2 \sin^2 \varphi + m_{\hpi}^2 \cos^2\varphi\,,\notag\\
	\mbox{and} \qquad \lambda v^2 = \sqrt{2} \sin2\varphi (m_{\hpi}^2 - m_{\lpi}^2)\,. \label{eqn:UPMS}
\end{gather}
Requiring the mixing operator~\eqref{eq:quartics} to be perturbative, and anticipating the possible $\lambda$ values from UV completions of the triplet-singlet framework, hereafter we shall generally require $|\lambda| \lesssim 2$. This in turn constrains the mixing angle $\varphi$ for a given mass splitting $m_{\hpi}^2 - m_{\lpi}^2$ and vice versa.

\subsection{Electroweak precision constraints}
\begin{figure}[t]
\includegraphics[width=0.45\textwidth]{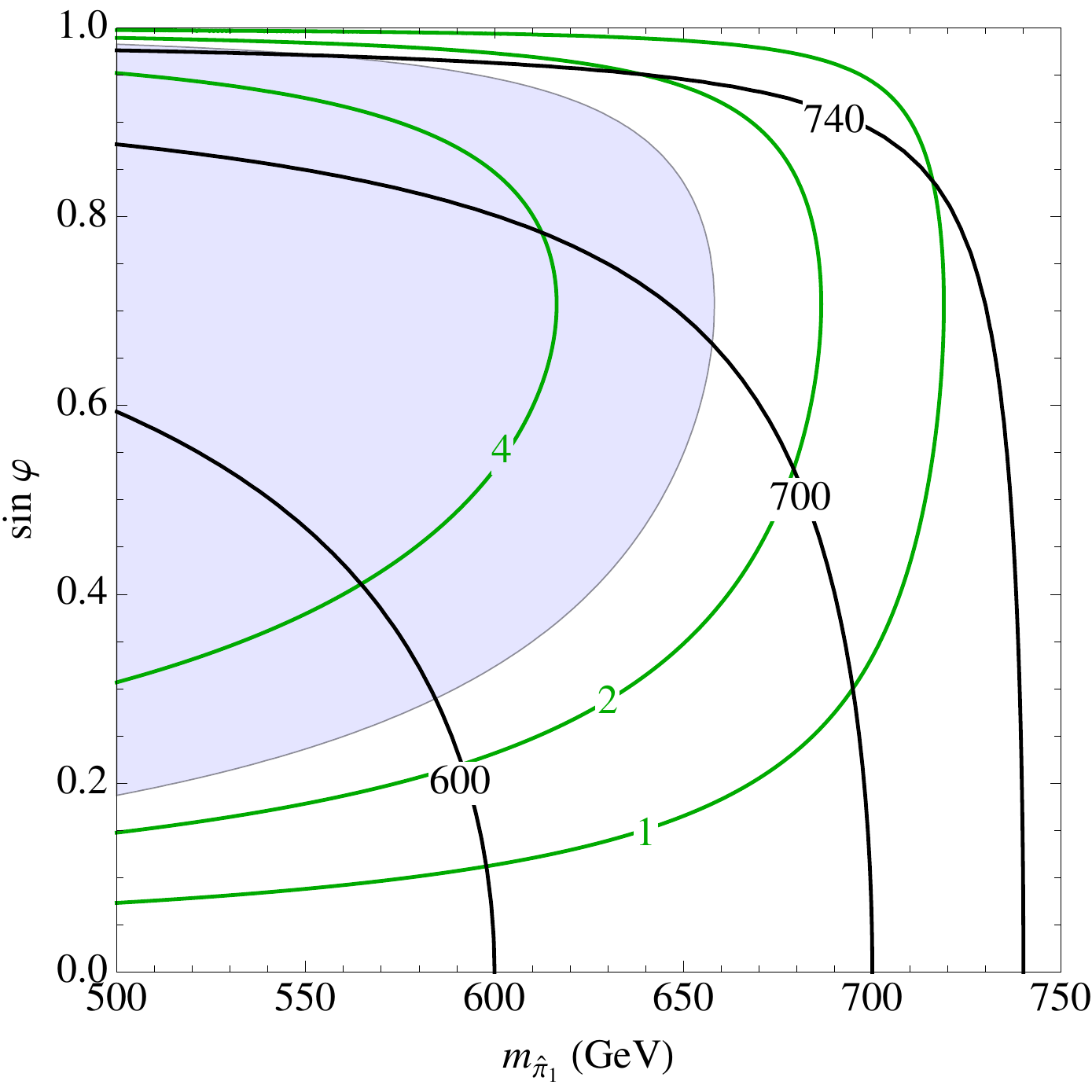}\hfill
\includegraphics[width=0.45\textwidth]{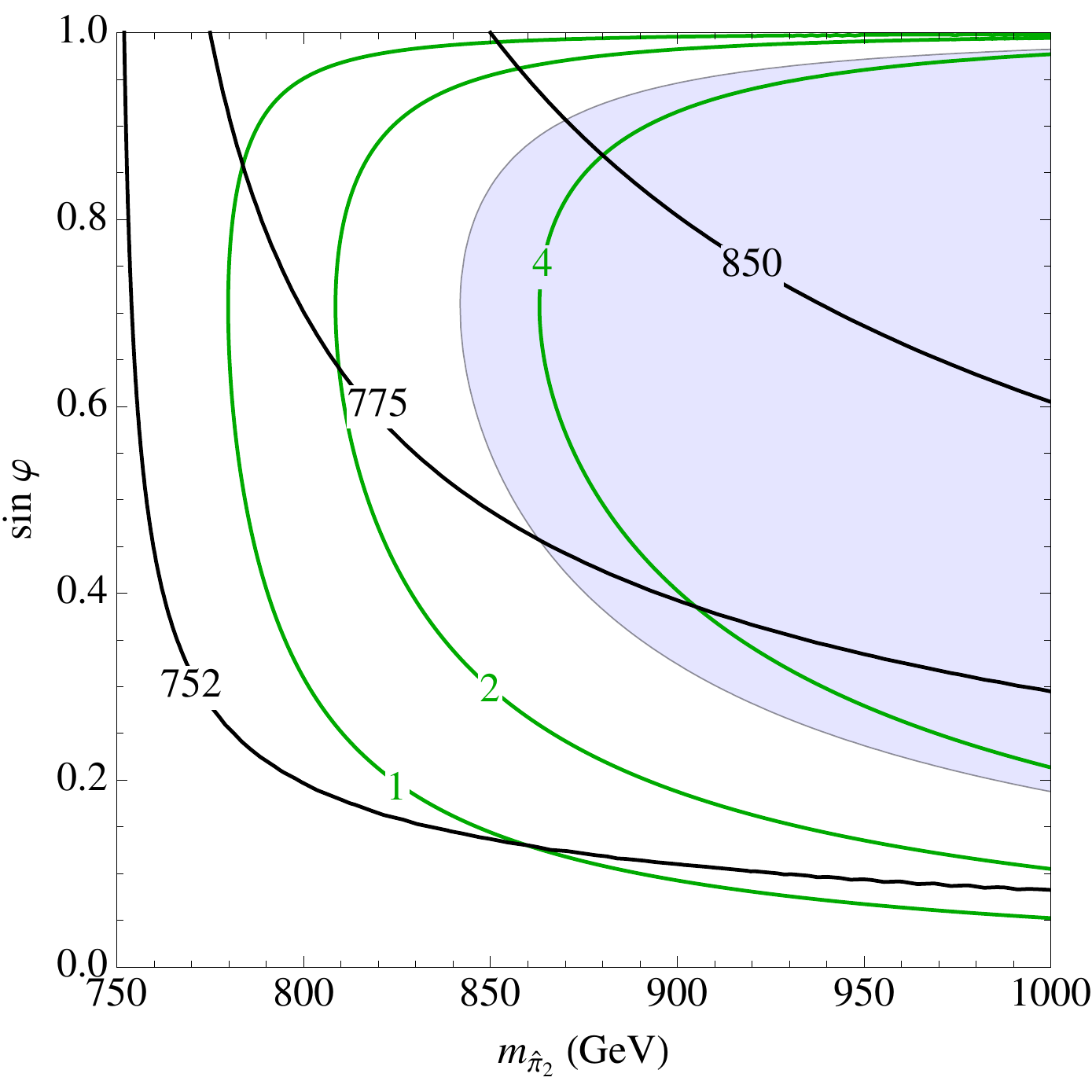}
\caption{Exclusion regions (light blue) from electroweak precision observables in the $\sin\varphi$--$m_{\lpi}$ plane (left) and $\sin\varphi$--$m_{\hpi}$ plane (right) for $m_{\hpi}=750$\,GeV and $m_{\lpi}=750$\,GeV, respectively.
Also shown are contours for $m_{\tplus}$ (black, in GeV) and $|\lambda|$ (green).} 
\label{fig:paramspace}
\end{figure}

The $H^\dagger \tPi H \szero$ operator in eq.~\eqref{eq:quartics} explicitly breaks custodial symmetry, and hence generates a one-loop contribution to the T-parameter from the operator
\begin{equation}
	\label{eqn:DTE}
	\mathcal{O}_T =  \frac{c_T}{2}\frac{\lambda^2}{16\pi^2} H^\dagger D_\mu H\;H^\dagger D^\mu H\,,
\end{equation}
in which one finds
\begin{equation}
\label{eq:Tparameter}
c_T= - \frac{\cos^22\varphi}{(m_{\hpi}^2-m_{\lpi}^2)^3}\bigg\{m_{\hpi}^4-m_{\lpi}^4+2m_{\lpi}^2m_{\hpi}^2\log\bigg[\frac{m_{\lpi}^2}{m_{\hpi}^2}\bigg]\bigg\} - \frac{1}{6}\sin^22\varphi \bigg(\frac{1}{m_{\lpi}^2}+\frac{1}{m_{\hpi}^2}\bigg)\,.
\end{equation} 
Comparing eqs.~\eqref{eqn:UPMS} and \eqref{eqn:DTE}, we see that in order to keep $\lambda$ small, and hence constrain T-parameter shifts, $\Delta T$, one requires $\sin 2\varphi$ to vanish as the splitting $m_{\hpi}^2 - m_{\lpi}^2$ grows large. Conversely, to maintain an $\mathcal{O}(1)$ mixing, the upper bound on $\Delta T$ requires an upper bound on $m_{\hpi}^2 - m_{\lpi}^2$. Fixing $m_{\lpi} = 750$\,GeV ($m_{\hpi} = 750$\,GeV), we show the allowed $m_{\hpi}$\!--$\sin\varphi$ ($m_{\lpi}$\!--$\sin\varphi$) parameter space in Fig.~\ref{fig:paramspace}, applying the $2\sigma$ electroweak precision (EWPT) bound \cite{Chankowski:2006hs,Chen:2013kfa,Agashe:2014kda},
\begin{equation}
	\frac{\sqrt{c_T}\lambda v^2}{16 \pi} \lesssim 3\,\mbox{GeV}\,,
\end{equation}
corresponding to $\delta \rho \lesssim 6\times10^{-4}$. Also shown are contours of $m_{\tpm}$ and $\lambda$, as determined by eq.~\eqref{eqn:UPMS}. 

As expected, we see in Fig.~\ref{fig:paramspace} that for small and large $\sin\varphi$, the splitting $m_{\hpi}^2 - m_{\lpi}^2$ may become arbitrarily large, but is bounded by EWPT constraints if the mixing is large. In the region allowed by EWPT we find roughly $|\lambda| \lesssim 2$, consistent with perturbativity of the effective theory. In this region, the $\lhpi$ splitting is at most 60 GeV in the maximally mixed case.

We also note that because $\lambda$ only couples the Higgs to neutral states, the $h\rightarrow \gamma\gamma / \gamma Z$ rates are not directly modified at 1-loop; wave-function renormalization is typically the dominant effect on Higgs couplings, and could potentially lead to $\sim0.1\%$--$1\%$ level modifications of Higgs couplings for larger values of $\lambda$ \cite{Craig:2013xia,Englert:2013tya}.

\subsection{Pair production}
The $\tzero$ and $\tpm$ states can be pair produced through the electroweak Drell-Yan process $q\bar{q} \to W^* \to \tpm \tzero$ or $q\bar{q} \to Z^*/\gamma^* \to \tplus \tminus$. Although the cross sections for these processes are small, they yield spectacular signatures comprising double diboson resonances in the final state. The parton level Drell-Yan cross-sections are
\begin{align}
		\hat{\sigma}_{\tzero\tpm} & = \frac{e^4}{96 \pi s_W^4} \frac{\hat s}{(\hat s - m_W^2)^2}\bigg[\bigg(1 - \frac{(m_{\tzero}+m_\tpm)^2}{\hat s}\bigg)\bigg(1 - \frac{(m_{\tzero}-m_\tpm)^2}{\hat s}\bigg)\bigg]^{3/2}\,\\
		\hat{\sigma}_{\tplus\tminus} & = \sum_{Q}\frac{e^4}{48 \pi s_W^4} \frac{\hat s}{(\hat s - m_Z^2)^2}\bigg[1 - \frac{4m_{\tplus}^2}{\hat s}\bigg]^{3/2}\\
		& \qquad\qquad \times \Big\{\Big[a_L + Qs_W^2\big(1 - m_Z^2/\hat s\big)\Big]^2 + \Big[a_R + Qs_W^2\big(1 - m_Z^2/\hat s\big)\Big]^2\Big\}\,,\notag
\end{align}
in which $a_L\equiv Q c_W^2-1/6$, $a_R\equiv -Q s_W^2$, and $Q=2/3$ or $-1/3$ is the electric charge of the initial state up or down quarks. Including triplet-singlet mixing, the mass eigenstate cross-sections $\hat{\sigma}_{\lpi\tpm} = \cos^2\varphi\,\hat{\sigma}_{\tzero\tpm}|_{m_\tzero \to m_\lpi}$ and $\hat{\sigma}_{\hpi\tpm} = \sin^2\varphi\,\hat{\sigma}_{\tzero\tpm}|_{m_\tzero \to m_\hpi}$. For the $\lpi\tpm$ modes, the $\lpi$ branching ratios depend on the underlying parameters of the model, as in eqs.~\eqref{eqn:PW} and \eqref{eqn:SPW}. We will discuss this in detail in Sections \ref{sec:diphoton} and \ref{sec:BRR}. It is therefore possible that the $\lpi$ decays mostly to jets, rather than EW gauge bosons. In contrast, the rate for the $\tplus\tminus$ mode is completely fixed up to the mass of the triplet, $m_{\tPi} = m_{\tplus}$. Also the $\tpm \to WZ$, $W\gamma$ branching ratios are fully determined up to small phase space effects: $\Br^{\tpm}_{W\g} = c_W^2\simeq 0.8$ and $\Br^{\tpm}_{WZ} = s_W^2\simeq 0.2$; the $\tpm \to \pi_1 W^{\pm*}$ branching ratios to three-body final states are comparatively negligible due to a strong virtuality penalty.  This pair production mode therefore provides a robust test of the triplet-singlet framework. 

The corresponding Drell-Yan pair production rates for the 13 TeV LHC are shown in Table~\ref{tab:pairproductionrates} for the case where $\lpi$ is close to a pure triplet, $\cos^2\varphi \simeq 1$.  At present, the most sensitive probe for $\lpi\tpm$ pair production is the search for three photons \cite{Aad:2015bua}. This search has a reach of several fb, but is currently not optimized for the particular signature at hand. A more optimal set of cuts was proposed in Refs.~\cite{Kilic:2010et,Freitas:2010ht}, and it should be possible to probe the $\lpi \tplus \to \g\g W\g$ mode with more data,  provided that $\lpi$ has a sufficiently large branching ratio to photons. 

While the $ \tplus \tminus \to W^+\g W^-\g$ rate is fixed by the SM electroweak couplings, and is therefore always undiluted, it is also experimentally more challenging because of the combinatorial background and the relatively small branching ratios of the leptonic modes. The search for $W\g$ resonances \cite{Aad:2014fha} is a priori relevant for this scenario, but currently sets only constraints in the $\sim 10$ fb regime and therefore will not be sensitive to this pair production signal for the projected LHC luminosities. With enough data it may nevertheless be possible to probe the $ \tplus \tminus$ pair production via a dedicated analysis that makes use of the full structure of the event, for example by requiring two hard photons and at least one lepton.

Finally, neutral pair production may proceed via gluon fusion through the Higgs portal generated by the operators \eqref{eq:quartics}, with parton level cross-section 
\begin{equation}
	\label{eqn:HPPP}
	\hat{\sigma}_{\pi^0 \eta} \simeq K\left(\frac{\alpha_s}{4\pi}\right)^2 \frac{\lambda^2}{32 \pi \hat{s}} |\mathcal{A}^h_{1/2}(4m_t^2/\hat{s})|^2\,,
\end{equation}	
in which $|\mathcal{A}^h_{1/2}(4m_t^2/\hat{s})| \simeq 0.6$ is the top loop function.  Rates for the mass eigenstates themselves may be obtained by including the appropriate mixing angle and symmetry factors, and for charged states under the replacement $\lambda \to \lambda_\tPi$. For the 13 TeV LHC, the cross-section~\eqref{eqn:HPPP} corresponds to a small production rate $\sim 0.03 \lambda^2$\,fb, which will likely not be detectable unless $\lambda \sim 2$ and the branching ratio to $4\g$ is $\mathcal{O}(1)$. The latter can occur if the single production occurs through photon fusion, see section \ref{sec:photonfusion}.

\begin{table}[t]
\newcolumntype{C}[1]{>{\centering\arraybackslash}  m{#1} <{} }
\begin{tabular}{|C{1.1cm}|C{2cm}|C{1.5cm}|C{1.1cm}|C{2cm}|C{1.5cm}|C{1.1cm}|C{2cm}|C{1.5cm}|}\hline
	Mode & Final State & $\sigma$ (fb) & Mode & Final State & $\sigma$ (fb) &Mode & Final State & $\sigma$ (fb)\\
	\hline
	\multirow{4}{*}{\rotatebox[origin=c]{0}{$\lpi \tplus$}}		& $VV W^+\g$ 		& 0.55 	& \multirow{4}{*}{\rotatebox[origin=c]{0}{$\lpi \tminus$}} 	&$VV W^-\g$		&0.19	& \multirow{4}{*}{\rotatebox[origin=c]{0}{ $\tplus\tminus$}} 	&$W^+\g W^-\g$	&0.27 \\
												&$VV W^+Z$		&0.16	&											&$VV W^-Z$		&0.06		&				
				&$W^+Z W^-\g$	&0.16\\	
	&&&&&&		&$W^+Z W^-Z$	&0.02\\
	\cline{2-3}\cline{5-6}\cline{8-9}
	&Total&0.72&&Total&0.24&&Total&0.45\\
	\hline
\end{tabular}
\caption{Pair production rates in fb for $m_\lpi=m_\tpm$=750 GeV, obtained with MSTW 2008 pdf's \cite{Martin:2009iq}. Here $VV$ stands for the sum over all final states $\lpi \to \g\g$, $ZZ$, $Z\g$, $WW$ and $gg$.}
\label{tab:pairproductionrates}
\end{table}

\section{Diboson phenomenology}
\label{sec:diphoton}
We now proceed to examine the diboson signatures produced by either $\lpi$ or $\hpi$ or both. This phenomenology is sensitive to various mass splitting thresholds and mixing regimes that we examine in turn. Throughout this analysis we apply a narrow width approximation to the $\lhpi$ decay rates, and assume gluon fusion production, such that the $pp \to \lhpi \to VV$ rate
\begin{equation}
	R^{\lhpi}_{VV}=K\frac{\pi^2}{8}\frac{\Gamma^{\lhpi}_{gg}}{m_{\lhpi}}\mathcal{L}_{gg}(m_{\lhpi})\text{Br}^{\lhpi}_{VV}\,,
\end{equation}
in which $VV = \g\g$, $Z\g$, $ZZ$ or $WW$. Here $\mathcal{L}_{gg}$ is the gluon luminosity function \cite{Martin:2009iq} -- $\mathcal{L}_{gg} \simeq 3850$\,pb when evaluated at 750 GeV -- and we include an estimated NNLO K-factor of $K\simeq3$ \cite{Harlander:2002vv}.\footnote{Ref.~\cite{Harlander:2002vv} calculates the NLO and NNLO K-factors for pseudoscalar production in the infinite top quark mass limit, which is equivalent to the effective operator generated by the anomaly.} Since the true diphoton rate, if non-zero, is still poorly known, we take 5\,fb as our benchmark value hereafter. 

For concreteness, in this section we consider two coupling benchmarks: 
\begin{equation}
\label{eqn:BD}
\begin{split}
	\text{A:} \qquad & c_{\tPi}=5\,,\qquad c_{1}=1\,,\qquad c_{2}=2\,,\qquad c_{3}=-\frac{5}{4}\,;\\
	\text{B:} \qquad & c_{\tPi} = c_3 = 5\,, \qquad c_{1,2} = 0\,.
\end{split}
\end{equation}
Benchmark A anticipates values predicted by an $SU(N_c = 5)$ composite model that we present in Section \ref{sec:composite} below. Benchmark B encodes an instructive toy theory in which the singlet $\szero$ is coupled only to QCD. Such a theory can be achieved in an \emph{ad hoc} perturbative UV completion of the triplet-singlet framework. Note that for benchmark B, the $\lhpi \to WW$ channel vanishes. Moreover, for any choice of couplings it is always the case that
\begin{equation}
	\Gamma^{\lpi}_{gg}=\sin^2\varphi\, \Gamma^{\szero}_{gg}\,, \qquad \mbox{and} \qquad \Gamma^{\hpi}_{gg}=\cos^2\varphi\, \Gamma^{\szero}_{gg}\,.
\end{equation}
For numerical evaluations in this section, we take $\alpha_s(m_{\lhpi}) \simeq 0.09$, estimated at one-loop order.

\subsection{Unresolved resonances: \texorpdfstring{$m_{\hpi}-m_{\lpi}<40$}{40}\,GeV }
If the splitting between both states is smaller than the experimental resolution of roughly 40 GeV, both diphoton resonances $\lhpi \to \g\g$ are misidentified as a single broad resonance.  (If a signal is observed in a higher mass resolution channel, e.g. $ZZ\rightarrow 4l$, the presence of two resonances may nevertheless be resolved.) For such a small mass splitting, $|\lambda|\lesssim1.3$ regardless the value of $\sin\varphi$, and we therefore do not need to restrict the range of the mixing angle. Assuming still $\Delta m_{\lhpi}^2 \gg m_{\lhpi} \Gamma_{\lhpi}$, so that interference effects may be neglected -- a safe assumption since from eqs~\eqref{eqn:PW}, $\Gamma_{\lhpi} \lesssim 10$~MeV for either benchmark -- the effective observed rate in $\g\g$ is then
\begin{equation}
	R^{\text{eff}}_{\g\g} \equiv R_{\g\g}^{\lpi} + R_{\g\g}^{\hpi} = K\frac{\pi^2}{8}\left[\frac{\Gamma^{\lpi}_{gg}}{m_{\lpi}}\mathcal{L}_{gg}(m_{\lpi})\text{Br}^{\lpi}_{\g\g}+\frac{\Gamma^{\hpi}_{gg}}{m_{\hpi}}\mathcal{L}_{gg}(m_{\hpi})	\text{Br}^{\hpi}_{\g\g}\right],\\
\end{equation} 
and similarly for the other decay channels. In the approximation that $ m_{\lpi}\simeq m_{\hpi}$, the effective diboson rates ratios
\begin{equation}
	\label{eqn:URR}
	\frac{R^{\text{eff}}_{VV}}{R^{\text{eff}}_{\g\g}} \simeq \frac{\Br^{\lpi}_{VV} + \cot^2\varphi \Br^{\hpi}_{VV}}{\Br^{\lpi}_{\g\g} + \cot^2\varphi \Br^{\hpi}_{\g\g}}\,,
\end{equation}
in which the branching ratios can be directly computed from eqs.~\eqref{eqn:TGI}--\eqref{eqn:SPW} and \eqref{eqn:MSR}.

\begin{figure}[t]
\includegraphics[width=0.47\textwidth]{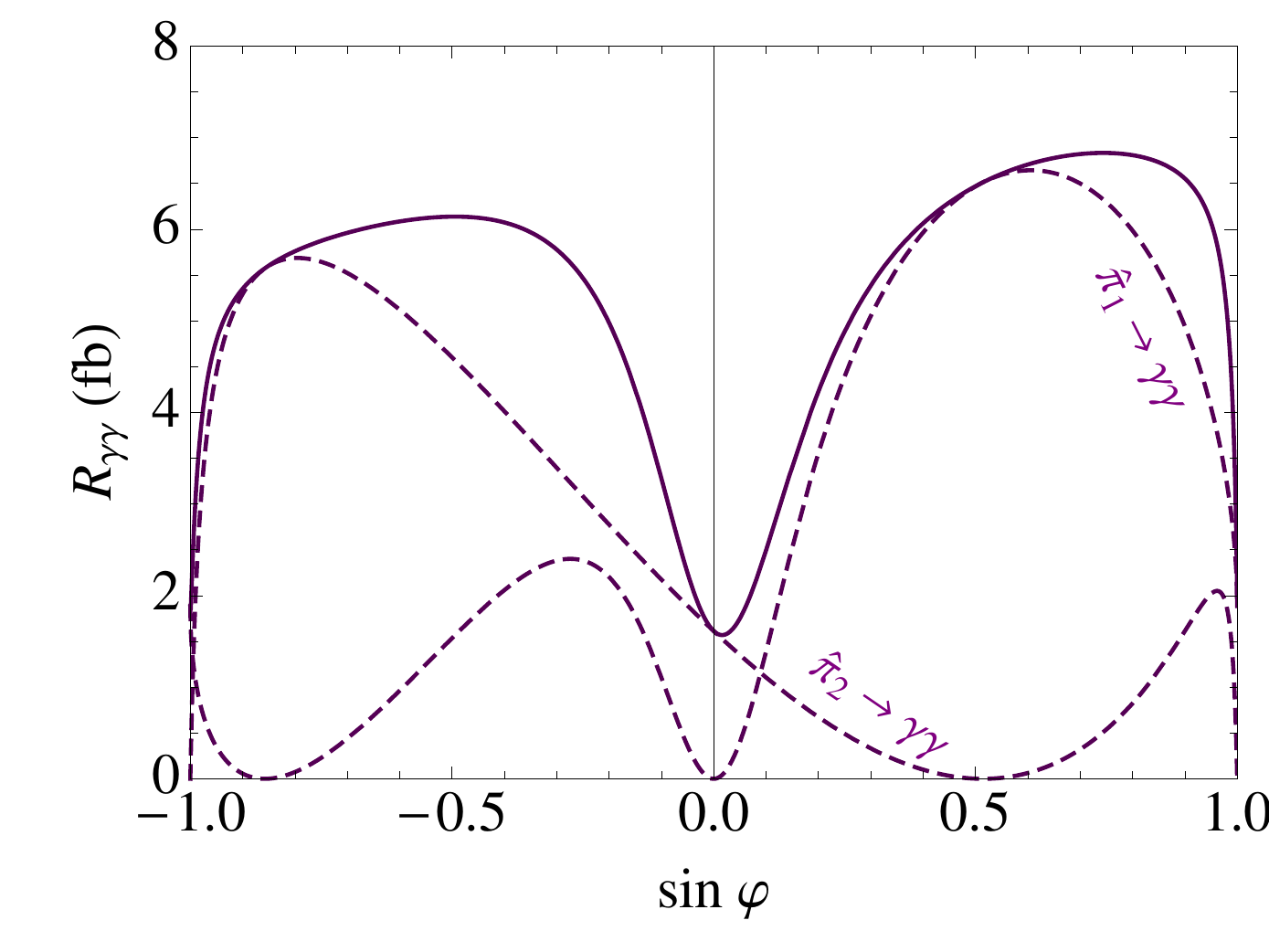}\hfill
\includegraphics[width=0.47\textwidth]{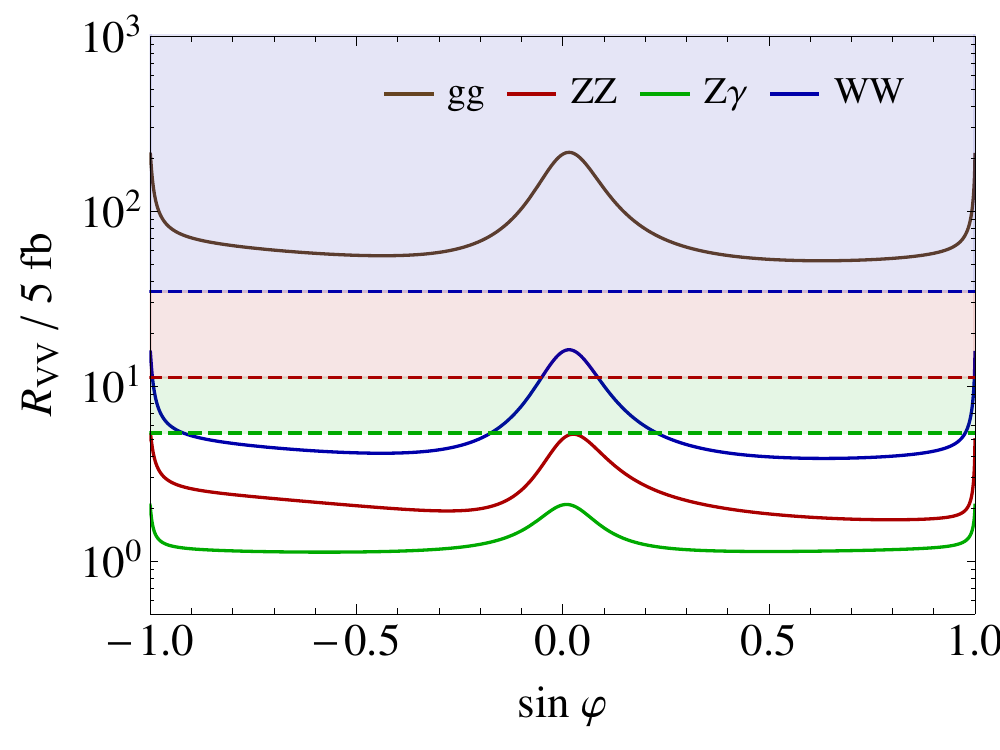}
\caption{Left: Diphoton rates $R^{\lhpi}_{\g\g}$  (dashed purple) and total effective rate $R^{\text{eff}}_{\g\g}$ (solid purple) as a function of $\sin\varphi$ for the unresolved resonances scenario at benchmark A, with $m_{\lpi}= 730$\,GeV, $m_{\hpi}= 770$\,GeV and $f$=1 TeV. Right: Current bounds on each diboson channel, normalized by the diphoton rate, fixing $R^{\text{eff}}_{\g\g}=5$ fb. For each diboson channel, the corresponding exclusion region is displayed. Exclusion regions apply only to curves of the same color, bounds from Table~\ref{tab:bounds}.}
\label{fig:C1}
\end{figure}

In Fig.~\ref{fig:C1} we show the effective diphoton rate as a function of the mixing angle for $m_{\hpi}-m_{\lpi}\simeq 40$ GeV and $f=1$\,TeV, as well as the current constraints on the ratios $R^{\text{eff}}_{VV}/R^{\text{eff}}_{\g\g}$, for benchmark A \eqref{eqn:BD}, with $m_{\lpi}=730$\,GeV, $m_{\hpi}=770$\,GeV. (The masses are chosen such that the effective resonance is centred at an invariant mass of $750$\,GeV.) In the right hand panel of Fig.~\ref{fig:C1} we vary $f$ such that $R^{\text{eff}}_{\g\g}=5$ fb remains fixed. The bounds used in these and subsequent figures are summarized in Table~\ref{tab:bounds}.

\begin{table}[t]
\begin{tabular}{|C{3cm}| C{5cm} | C{5cm}|}
	\hline
	Mode (VV) &750 GeV&800 GeV\\ \hline\hline
	$\g\g$	&--&1.1~\cite{CMS-PAS-EXO-12-045}\\
	$Z\g$	&5.4~\cite{ATLAS-CONF-2016-010}&8.0~\cite{ATLAS-CONF-2016-010}\\
	$ZZ$		&11.3~\cite{Aad:2015kna}&11.3~\cite{Aad:2015kna}\\
	$WW$	&34.8~\cite{Aad:2015agg}&33~\cite{Aad:2015agg}\\
	$gg$		&2000~\cite{CMS:2015neg}&1900~\cite{CMS:2015neg}\\
	\hline
\end{tabular}
\caption{Bounds on the rates for diboson resonances with mass of 750 GeV and 800 GeV, normalized against the estimated rate of the diphoton excess at 750 GeV, $R_{\g\g} =5$\,fb. \label{tab:bounds}}
\end{table}

We see from Fig.~\ref{fig:C1} that an $\mathcal{O}(1)$ mixing angle may produce the claimed $\sim5$\,fb diphoton rate, while all other diboson rates are simultaneously consistent with current constraints. If the rate from one of the resonances drops well below the other  -- e.g. near $\sin\varphi \simeq 0.5$ or $\sin\varphi \simeq -0.9$ --  the interpretation of the diphoton signal as a set of overlapping resonances forming a single, broad resonance is lost. 

For benchmark B, the ratios \eqref{eqn:URR} can be expressed explicitly,
\begin{gather}
	\label{eqn:BC1}
  \frac{R^{\text{eff}}_{WW}}{R^{\text{eff}}_{\g\g}} = 0\,,\qquad \frac{R^{\text{eff}}_{ZZ}}{R^{\text{eff}}_{\g\g}} = 1\,, \qquad \frac{R^{\text{eff}}_{Z\g}}{R^{\text{eff}}_{\g\g}} = 2\cot^2(2\theta_W)\simeq 0.8\,,\\ 
  \mbox{and}\qquad \frac{R^{\text{eff}}_{gg}}{R^{\text{eff}}_{\g\g}} = \frac{2 \mathcal{C}}{s^2_{2W}}\frac{ 4(\sin^6\varphi + \cos^6\varphi) + \mathcal{C}\sin^2 2\varphi}{(1 + \mathcal{C})\sin^2 2\varphi}\,,
\end{gather}
in which 
\begin{equation}
	\mathcal{C} \equiv \frac{\Gamma^{\tzero}_{gg}}{\sum_{VV}\Gamma^{\tzero}_{VV}} = 4(c_3/c_{\tPi})^2(\alpha_s/\alpha)^2 \sin^2 2\theta_W \simeq 430\,. 
\end{equation}	
Since $\mathcal{C} \gg 1$, the digluon rate is mostly flat, except for $\varphi \to 0$ or $\pm\pi/2$, and one finds in this flat region $R^{\text{eff}}_{gg}/R^{\text{eff}}_{\g\g} \simeq 76$ for benchmark B. The effective diphoton rate itself
\begin{equation}
	\label{eqn:C1DR}
		R^{\text{eff}}_{\g\g} \simeq K \frac{\pi^2}{8}\frac{ \Gamma^{\tzero}_{\g\g}}{m_{\lpi}} \mathcal{L}(m_{\lpi})_{gg} c^2_{\tPi} \simeq 2.5\,\text{fb} \bigg[\frac{1.5\,\TeV}{f}\bigg]^2(c_{\tPi}/5)^2 \,, 
\end{equation}
up to $\mathcal{O}(1/\mathcal{C})$ corrections. Comparing the bounds from Tab.~\ref{tab:bounds} with eqs.~\eqref{eqn:BC1}, one sees that similarly all other diboson rates are consistent with current constraints.

\subsection{Resolved resonances: \texorpdfstring{$40\,\GeV <m_{\hpi}-m_{\lpi}<m_W$}{40mw}}

For mass splittings above $40$\,GeV, the $\lpi \to \g\g$ and $\hpi \to \g\g$ resonances may be resolved by experiments. We identify the $\lpi$ state as the $750$\,GeV diphoton resonance and we require $R^{\lpi}_{\g\g} \simeq 5$\,fb, while all other rates are subject to experimental constraints.  In this regime, since $m_{\hpi}-m_{\lpi}<m_W$, the $\hpi \to \tpm W^\ast$ decay only proceeds off-shell. For instance, in the $m_{\hpi}-m_{\lpi}\ll m_W$ limit, the rate is
\begin{equation}
	\G^{\hpi}_{\tpm W^{\ast}} \simeq \sin^2\varphi \frac{\alpha^2}{15 \pi s_W^4}\frac{(m_{\hpi}-m_{\tpm})^5}{m_W^4}\,.
\end{equation}
For both our benchmarks in the regime $m_{\hpi} < 800$\,GeV, the branching ratio for this process never exceeds $4\times 10^{-4}$, and we hereafter neglect this channel. Further for $m_{\hpi} < 800$\,GeV, $|\lambda|\lesssim1.7\sin2\varphi$, which is mildly large for maximal mixing, but still perturbative.

\begin{figure}[t]\centering
\includegraphics[width=0.5\textwidth]{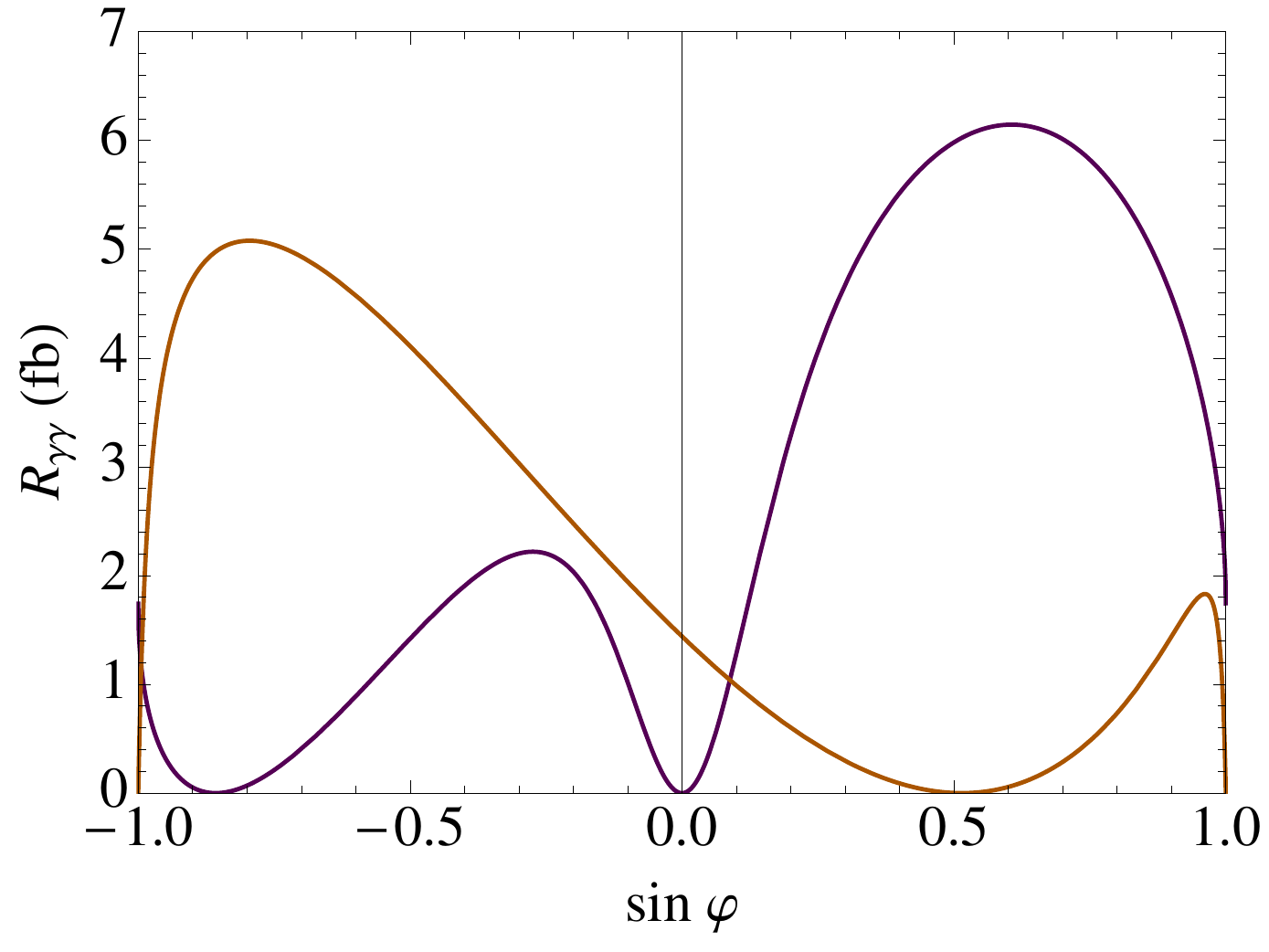}
\caption{Left: Diphoton rates $R^{\lpi}_{\g\g}$  (purple) and $R^{\hpi}_{\g\g}$ (orange) for the split resonances scenario at benchmark A. Here $m_{\lpi}= 750$\,GeV, $m_{\hpi}= 800$\,GeV, and $f=1$\,TeV.}
\label{fig:2Arate}
\end{figure}

\label{sec:DE}
\begin{figure}[t]
\includegraphics[width=0.5\textwidth]{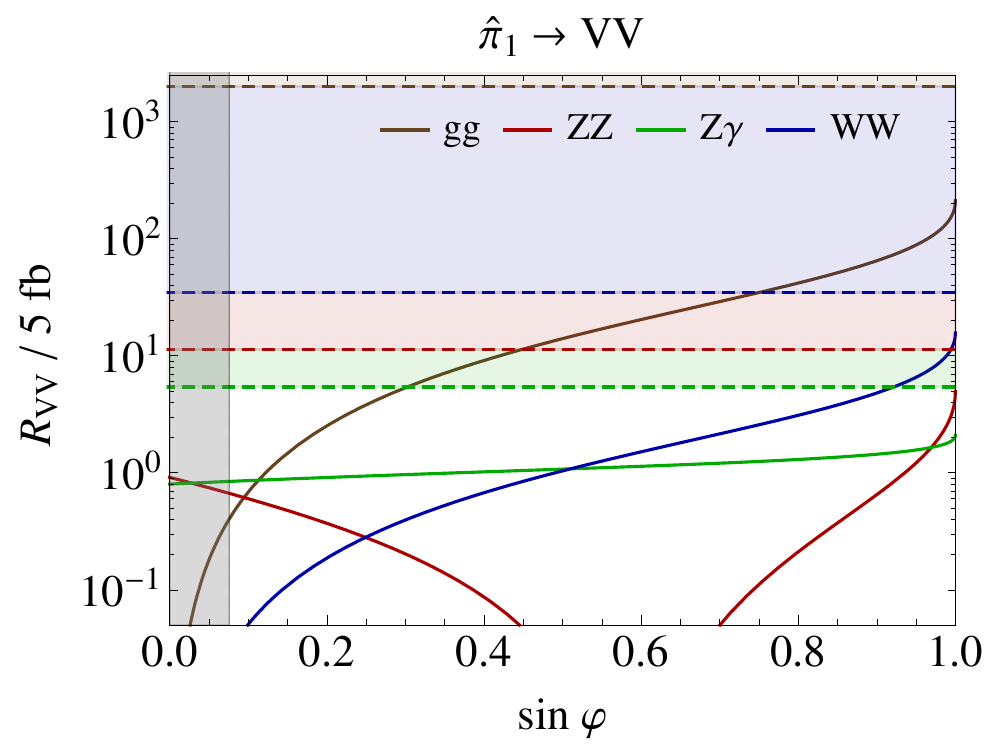}\hfill
\includegraphics[width=0.5\textwidth]{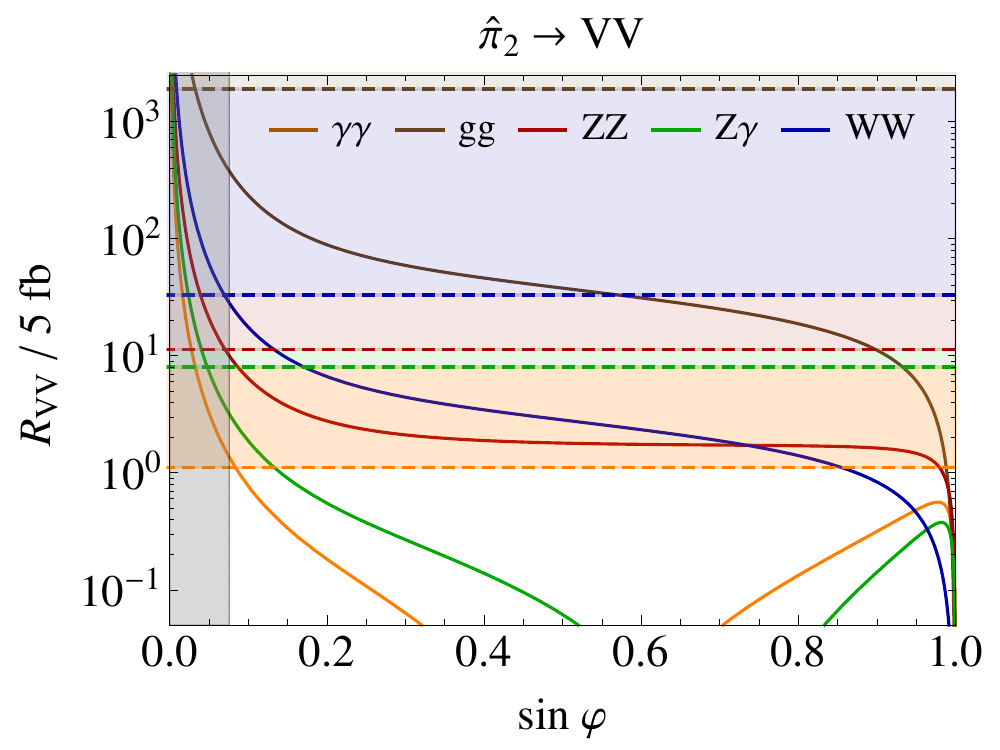}
\caption{Rates for each diboson channel, normalized by the diphoton rate, as a function of $\sin\varphi$, $m_{\lpi}= 750$\,GeV, $m_{\hpi}= 800$\,GeV. For each diboson channel, the corresponding exclusion region is displayed, fixing $R^{\lpi}_{\g\g}=5$\,fb. Exclusion regions apply only to curves of the same color, bounds from Table~\ref{tab:bounds}. Vertical gray shading indicates $f<400$ GeV.}
\label{fig:C2}
\end{figure}

For benchmark A, we show in Fig.~\ref{fig:2Arate} the diphoton rate for both $\lpi$ and $\hpi$ with $m_{\hpi} = 800$\,GeV and $f = 1$\,TeV. For negative values of $\sin\varphi$ the diphoton rate from $\hpi$ exceeds the rate for $\lpi$, which is heavily disfavored by the data. We therefore do not consider this region further. In Fig.~\ref{fig:C2} we show the rates of the remaining channels, normalized against the $\lpi$ diphoton rate, where we again vary $f$ to keep $R^{\lpi}_{\g\g} = 5$\,fb fixed. For $f< 400$\,GeV we expect it to be challenging to UV complete the effective theory with the composite theories that we consider in section \ref{sec:composite}. This region is marked by the gray shading on Fig.~\ref{fig:C2}. The decay modes of $\lpi$ are unconstrained, but one requires $\sin\varphi\gtrsim 0.1$ in order to evade the bounds on $\hpi\rightarrow \g\g$ and $\hpi\rightarrow WW$.

For benchmark B in this regime, the decay of $\lpi$ occurs purely through the $c_{\tPi}$ or $c_3$ couplings. The decay rate ratios 
\begin{gather}
	\label{eqn:BC2}
	\frac{R^{\lpi}_{WW}}{R^{\lpi}_{\g\g}} = 0\,, \qquad\frac{R^{\lpi}_{ZZ}}{R^{\lpi}_{\g\g}} = 1\,, \qquad \frac{R^{\lpi}_{Z\g}}{R^{\lpi}_{\g\g}} = 2\cot^2(2\theta_W)\simeq0.8\,\\
	\mbox{and} \qquad  \frac{R^{\lpi}_{gg}}{R^{\lpi}_{\g\g}} = \frac{2 \mathcal{C}\tan^2\varphi}{s^2_{2W}}\simeq 1200 \tan^2\varphi\,,
\end{gather}
and the diphoton rate itself 
\begin{equation}
	\label{eqn:C2DR}
	R^{\lpi}_{\g\g}  \simeq K \frac{\pi^2}{8}\frac{ \Gamma^{\lpi}_{\g\g}}{m_{\lpi}} \mathcal{L}_{gg} (m_{\lpi})c^2_{\tPi}   \frac{\mathcal{C}\sin^2\varphi}{1+ \mathcal{C}\tan^2\varphi} \simeq 3.3\,\text{fb} \bigg[\frac{1200\,\GeV}{f}\bigg]^2(c_{\tPi}/5)^2 \cos^2\varphi\,, 
\end{equation}
provided $\mathcal{C}\tan^2\varphi \gg 1$. In practice this approximation holds if $\tan\varphi \gtrsim 0.15$. Comparing the bounds from Tab.~\ref{tab:bounds} with eqs.~\eqref{eqn:BC2}, one sees that the diboson rates are well within current bounds. The digluon rate bound requires that $\tan \varphi \lesssim 1.3$, which is easily satisfied over most of the $\sin\varphi$ parameter space. 

The corresponding $R^\hpi_{VV}$ rates are obtained from eqs.~\eqref{eqn:BC2} and \eqref{eqn:C2DR} under the exchange $\sin \varphi \leftrightarrow \cos\varphi$. The bounds on these rates are easily evaded for small or $\mathcal{O}(1)$ $\sin \varphi$, provided $\varphi$ is not so small that $R^{\hpi}_{gg}/R^{\lpi}_{\g\g}$ becomes larger than $\sim 10^3$. I.e. one requires only $\cot\varphi \lesssim 0.9$.

\subsection{Cascades: \texorpdfstring{$m_W < m_{\hpi}-m_{\lpi} < m_h$}{mwh}}
\label{sec:WS}
Next we consider the regime in which $m_W < m_{\hpi}-m_{\lpi} < m_h$. As soon as the mass splitting between $\hpi$ and $\tpm$ becomes larger than $m_W$, the tree-level channel $\hpi\rightarrow {\tpm W^\mp}$ opens up, with partial width
\begin{equation}
	\label{eqn:PWP}
	\Gamma^\hpi_{\tpm W^\mp } =\frac{\alpha}{s_W^2}\frac{m_{\hpi}^3\sin^2\varphi}{m_W^2}\bigg[1-\bigg(\frac{m_\tpm+m_W}{m_{\hpi}}\bigg)^2\bigg]^{3/2}\bigg[1-\bigg(\frac{m_\tpm-m_W}{m_{\hpi}}\bigg)^2\bigg]^{3/2}\,,
\end{equation}
summing over both final states. (Note the ${m_{\hpi}^2}/{m_W^2}$ enhancement for $\Gamma^\hpi_{\tpm W^\mp}$ from the Goldstone boson equivalence principle, that is unitarized in the large $m_\hpi$ limit via the relation $m_{\hpi}^2 \sin\varphi \sim \lambda v^2$, from eq.~\eqref{eqn:MSRD}.)

Since we wish to avoid non-perturbative values for $\lambda$, eq.~\eqref{eqn:UPMS} restricts us to the cases for which $\varphi$ is small, i.e. $\lpi$ and $\hpi$ are close to pure states.  This requires us to take $\sin\varphi\lesssim 0.2$. The dominant production channel is then of the singlet-like state, which we identify as the heavier $\hpi$, and the cascade $\hpi \to \tpm W^{\mp}$ becomes relevant. Note that for $\sin\varphi\lesssim$ 0.2, we have $m_{\tpm}-m_{\lpi} \lesssim 5$ GeV (see Fig.~\ref{fig:paramspace}).

The first possibility is that the $\lpi$ is the $750~\GeV$ state, while the gluon fusion production cross-section for the heavier $\hpi$ is much larger than for $\lpi$. Since $\Br^{\tpm}_{W^\pm\g} = c_W^2 \simeq 0.8$, the cascade $\hpi\rightarrow W^\pm \tpm \rightarrow W^\pm \g$ can have a large rate. This leads to a rather distinctive $WW\g$ final state, in which one of the $W$'s forms a resonance with the photon at $m_{\tpm}$. The strongest constraint on this process comes from the ATLAS search for $W\g$ resonances in the leptonic channel \cite{Aad:2013izg} . This search vetoes additional leptons, so we require the second $W$ to decay hadronically for an event to pass the cuts. We can then reinterpret this search as setting the bound $R^{\hpi}_{\tpm W^\mp}<29$\,fb for $m_{\tpm}=755$ GeV. To illustrate the strength of the constraints, we consider a mass benchmark point with $m_{\lpi}=750$ GeV and $m_{\hpi}=850$ GeV. In the mixing regime of interest -- $\sin\varphi\ll1$ -- the width of $\hpi$ is dominated by the dijet mode, which means that the dependence of the partial width to gluons drops out from the rate, giving
\begin{equation}
	R^\hpi_{\tpm W^\mp } \simeq K\frac{\pi^2}{8}\frac{\G^\hpi_{\tpm W^\mp }}{m_{\hpi}} \mathcal{L}_{gg}(m_{\hpi})< 29\;\text{fb}\,.
\end{equation}
To good approximation, the rate is therefore independent of $c_{1,2,3}$, $c_{\Pi}$ and $f$. From eq.~\eqref{eqn:PWP} this bound can therefore be re-expressed as a constraint on $\sin\varphi$,
\begin{equation}
	|\sin\varphi|\lesssim 0.022\,.
\end{equation}
It follows that the production cross-section for $\hpi$ is $\gtrsim 2000$ times larger than for $\lpi$.  For benchmark A this implies that $R^{\hpi}_{\g\g}/R^{\lpi}_{\g\g} \sim 15$. Requiring still that $R^{\lpi}_{\g\g} \simeq 5$\,fb, the consequent $\hpi \to \g\g$ rate is strongly excluded by current diphoton bounds \cite{CMS-PAS-EXO-12-045}. For benchmark B, one finds $R^{\hpi}_{gg}/R^{\lpi}_{\g\g} \sim 4700$, which is in tension with current bounds~\cite{CMS:2015neg}. In this latter case, we find the $750~\GeV$ diphoton rate can be accommodated with $f\lesssim 400\,\GeV$.

For benchmark A, a second possibility is that instead the singlet-like $\hpi$ is the $750~\GeV$ diphoton resonance, and the $\lpi$ is lighter than $670$\,GeV.  In this case the diboson phenomenology is similar to that of the pure singlet case, which has been studied in detail elsewhere~\cite{Backovic:2015fnp,Angelescu:2015uiz,Knapen:2015dap,Buttazzo:2015txu,Franceschini:2015kwy,Gupta:2015zzs,Agrawal:2015dbf,Aloni:2015mxa,Altmannshofer:2015xfo}. However,  the decays $\hpi\rightarrow W^\pm \tpm$ can give an unusual $WW\gamma/WWZ$ final state for the $750~\GeV$ resonance, with a rate much larger than the Drell-Yan pair production of $\tpm$. 
Current constraints from the $W\gamma$ search on this tree-level decay require $\sin\varphi\lesssim10^{-2}$; but a dedicated resonant tri-boson search might increase the reach.

\subsection{Cascades: \texorpdfstring{$m_{\hpi}-m_{\lpi} > m_h$}{mh}}
\label{sec:DCD}
In the mass splitting regime $m_{\hpi}-m_{\lpi} > m_h$, a second tree-level decay mode open up for $\hpi$, namely $\hpi \to \lpi h$. The partial width for $\hpi\rightarrow h\lpi$ is
\begin{equation}
\begin{split}
\Gamma^\hpi_{\lpi h} = &  \frac{\lambda^2v^2\cos^2 2\varphi}{16\pi m_{\hpi}}\bigg[1-\bigg(\frac{m_\lpi+m_h}{m_{\hpi}}\bigg)^2\bigg]^{1/2}\bigg[1-\bigg(\frac{m_\lpi-m_h}{m_{\hpi}}\bigg)^2\bigg]^{1/2}\,,
\end{split}
\end{equation}
where $\lambda$ can be expressed in terms of $\varphi$ and the mass splitting via eq.~\eqref{eqn:UPMS}. Since the $\lpi$ can subsequently decay to a diphoton final state, this raises the interesting possibility that the observed signal originates dominantly from a cascade decay, rather than from direct $\lpi$ production through gluon fusion.

\begin{figure}
\includegraphics[width=0.5\textwidth]{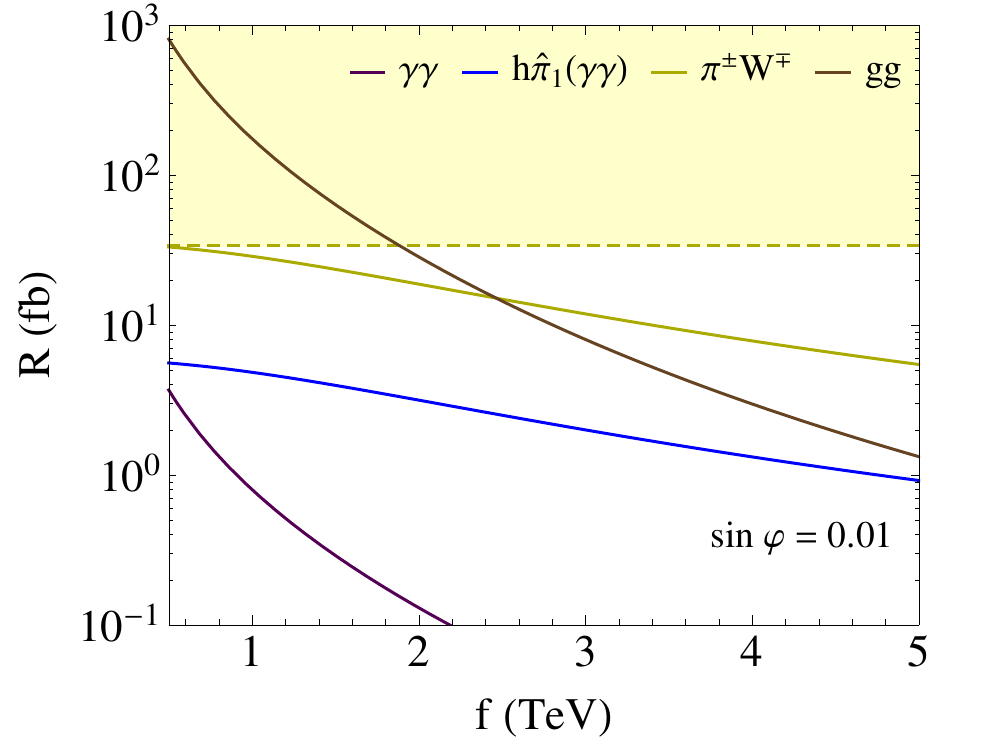}\hfill
\includegraphics[width=0.5\textwidth]{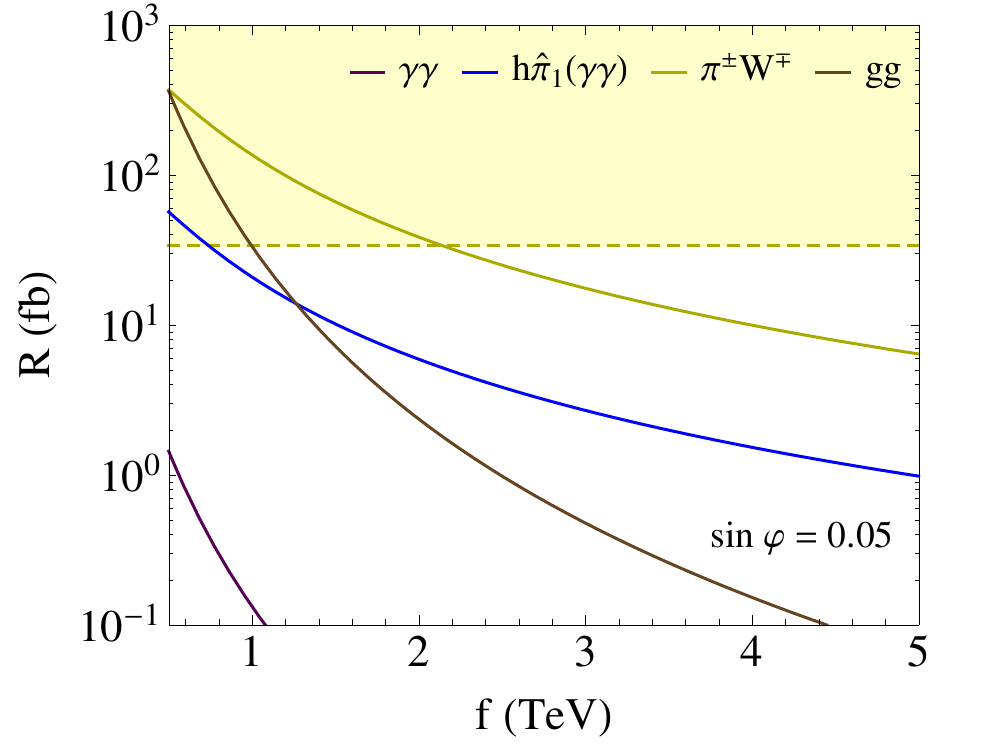}
\caption{Cascade diphoton rate for $\hpi \to h(\lpi \to \g\g)$ (blue) compared to the direct diphoton rate $R^{\hpi}_{\g\g}$ (purple) and the tree-level rate $R^{\hpi}_{\pi W}$ (yellow), at benchmark A with $m_\hpi = 900$\,GeV and $\sin\varphi = 0.01$ (right) and $0.05$ (left). Also shown is the exclusion region for $R^{\hpi}_{\pi W}$ (yellow shaded) from constraints on $W\g$ resonances (see Section~\ref{sec:WS}). Exclusion regions apply only to curves of the same color. }
\label{fig:heavysingletdecay}
\end{figure}

The absence of significant $p_T$ in the diphoton resonance data~\cite{moriond}, however, suggests that the $\hpi$ mass must be nearby the kinematic threshold for this cascade decay, so that the $h$ does not obtain a large transverse momentum. Further, production of a diphoton final state in association with a higgs also generically requires the presence of two $b$-jets. This is in tension with current jet counts for the diphoton data, disfavoring this method of producing the excess \cite{moriond}. However, should associated $b$-jets be observed in future data, in Fig.~\ref{fig:heavysingletdecay} we show the effective diphoton cross-section obtained from this cascade decay at benchmark A, with $m_{\hpi}=900$\,GeV. For small values of $f$ and small $\sin\varphi$ -- equivalently small $\lambda$ -- the decays $\hpi\rightarrow gg$ can dominate over the $\hpi \rightarrow \lpi h$ mode. For example, for $f\sim1$ -- $2$\,TeV, a $\sim 5$\,fb diphoton signal rate can be obtained with $\sin\varphi\simeq0.01$. Note that for these values of $\sin\varphi$ the rate for direct $\lpi\to\g\g$ production is negligibly small (see Fig.~\ref{fig:C2}) and the excess must therefore entirely come from the cascade decay. In this part of parameter space it is however possible have an $\mathcal{O}(1)$\,fb rate for $\hpi\rightarrow \gamma\gamma$. For larger values of $\sin\varphi$, the decays $\hpi \rightarrow \lpi h$ start to dominate over $\hpi \to gg$. For example, the signal can be fit with $f\sim3$ -- $5$\,TeV and $\sin\varphi\simeq0.05$. For even larger $\sin\varphi$, the benchmark is in tension with $W\g$ resonance constraints from the $\hpi\rightarrow \tpm W^\mp$ decay, as discussed in Section \ref{sec:WS}.

Again, for benchmark A there is another possibility that the resonance at 750 GeV is due to a mostly singlet $\hpi$, while $\lpi$ is lighter than $625$\,GeV and $\sin \varphi$ remains small. If $\sin\varphi\neq0$, the cascade decay modes $\hpi\to h\lpi$ and $\hpi\to \tpm W^\mp$ can be present, with $\lpi$ and $\tpm$ subsequently decaying to pairs of electroweak gauge bosons, as discussed above. These provide unusual 3-boson final states $W^\pm W^\pm \gamma / W^\pm W^\pm Z / hVV$ for the $750~\GeV$ resonance, and the cascade production can exceed the Drell-Yan rates for $\pi^\pm\pi^\pm$ and $\pi^\pm\lpi$ production.  Current constraints search on these tree-level decays require $\sin\varphi\lesssim10^{-3}$; a dedicated resonant tri-boson search might increase the reach.

\subsection{Photon fusion: \texorpdfstring{$m_{\hpi} \gg m_{\lpi}$}{pf} \label{sec:photonfusion} }
If the singlet state is decoupled or absent altogether, the production of the triplet must be achieved entirely from electroweak processes, which are dominantly photon-photon fusion \cite{Fichet:2015vvy,Csaki:2015vek,Csaki:2016raa,Harland-Lang:2016qjy}. This is in mild tension with 8 TeV LHC results, because the cross section only scales by a factor of three from 8 to 13 TeV, and requires a large partial width $\sim 50$\,MeV. The 13 TeV $\gamma\gamma$ rate for the benchmark model in this scenario is
\begin{equation}
R_{\gamma\gamma} = 0.1 {\rm~fb} \times \bigg[\frac{500\GeV}{f}\bigg]^2(c_{\tPi}/5)^2\,.
\end{equation}
Hence one requires a `t-Hooft coupling $c_{\tPi} \sim 35$, which is harder to achieve in simple UV completions, without requiring large numbers of flavors of exotic hypercharged states.

\section{Branching ratio relations}
\label{sec:BRR}

So far we have restricted ourselves to two benchmark examples with fixed anomaly coefficients, such that the branching ratios were only a function of $\sin\varphi$.  In general, however, the partial widths $\Gamma^{\lhpi}_{VV}$, $VV = \g\g$, $ZZ$, $Z\g$, $WW$, are generated by the three field strength operators~\eqref{eq:gaugebasis}, with three independent couplings $c_{\tPi}$, $c_1$ and $c_2$, in addition to the mixing parameter $\sin\varphi$. The pure triplet and pure singlet cases are generated by only one and two operators, respectively, which implies that the ratios of the partial widths live respectively in a zero and one dimensional parameter spaces. The mixed triplet-singlet framework, however, encodes both these pure regimes in its larger parameter space, and therefore admits a much greater flexibility of the relative diboson branching ratios. 

For the $\lpi$ alone resolved as the 750\,GeV resonance, the $\lpi$ branching ratio relations can be characterized in a two-dimensional parameter space. We show there are regions of parameter space in which no other $750~\GeV$ diboson modes will be observed even with $3000~\text{fb}^{-1}$ at LHC13. However, the second neutral resonance will have complementary branching ratios to the $750~\GeV$ state, and can generically also be discovered in diboson modes unless it has a very small singlet component. In the case that $\lhpi$ diphoton resonances are unresolved, the corresponding merged dibosons rates may also be described in a two-dimensional parameter space, giving different possibilities from a simple pure singlet or triplet.

\subsection{Resolved resonances}
\label{sec:BRRR}
If the $\lpi$ resonance can be resolved from the $\hpi$, the production terms drop out from the ratio of rates, and we can write $R^{\lpi}_{VV}/R^{\lpi}_{\g\g} = \Gamma^{\lpi}_{VV}/\Gamma^{\lpi}_{\g\g}$. This permits us to probe the underlying operator structure directly, as follows. For the pure triplet case, the relative $\lpi$ branching ratios are fixed by eqs.~\eqref{eqn:PTBR}, so that observation of any one of the ratios $R^{\lpi}_{VV}/R^{\lpi}_{\g\g}$ has the potential to exclude this scenario. Similarly, the pure singlet case can be potentially excluded with the observation of at least two ratios $R^{\lpi}_{VV}/R^{\lpi}_{\g\g}$ (see for instance Refs.~\cite{Craig:2015lra,Low:2015qho,Kamenik:2016tuv}). More generally, the structure of the triplet-singlet framework encoded in eqs.~\eqref{eqn:TGI}--\eqref{eqn:SPW} permits the three ratios $\Gamma^{\lpi}_{VV}/\Gamma^{\lpi}_{\g\g}$ to be expressed in terms of two polar parameters, $r>0$ and $\psi \in [0,2\pi)$, defined via
\begin{equation}
	\label{eq:polardef}
	r \cos\psi \equiv (c_1/c_{\tPi})\tan\varphi\,,\qquad r \sin\psi \equiv (c_2/c_{\tPi})\tan\varphi\,,\quad \text{so} \quad r = \bigg|\frac{\sqrt{c_1^2 + c_2^2}}{c_{\tPi}}\tan\varphi\bigg|\,.
\end{equation}
The radius $r$ interpolates between the zero dimensional pure triplet parameter space ($r=0$) and one-dimensional pure singlet parameter space ($r=\infty$). The ratio $\tan\psi = c_2/c_1$ controls the relative weight of the $B^{\mu\nu}\tilde B_{\mu\nu}$ and $W^{\mu\nu}\tilde W_{\mu\nu}$ operators. The electroweak diboson couplings of the $750~\GeV$ state become
\begin{equation}
\frac{\alpha c_{\tPi} \cos\varphi}{8\pi}  \frac{\lpi}{f} 
\left[
\frac{1}{c_W s_W} W^3_{\mu\nu}\tilde B^{\mu\nu} + 
\frac{r \cos\psi}{c_W^2} B_{\mu\nu}\tilde B^{\mu\nu} + 
\frac{r \sin\psi}{s_W^2}\left(W^3_{\mu\nu}{\tilde W^3}^{\mu\nu}+ 2W_{\mu\nu}^+{\tilde W^-}^{\mu\nu} \right)
\right]
\end{equation}
giving the relative branching ratios
\begin{equation}
	\label{eqn:PWRR}
	\begin{split}
	\frac{R^{\lpi}_{ZZ}}{R^{\lpi}_{\g\g}}  & = \Bigg(\frac{1-r \cos\psi \tan^2\theta_W - r\sin\psi \cot^2\theta_W}{1 + r \cos\psi + r\sin\psi}\Bigg)^2\,,\\
	\frac{R^{\lpi}_{Z\g}}{R^{\lpi}_{\g\g}}  & = \frac{1}{2}\Bigg(\frac{\cot\theta_W(2r\sin\psi + 1) - \tan\theta_W(2 r \cos\psi +1)}{1+ r \cos\psi + r\sin\psi}\Bigg)^2\,,\\
	\frac{R^{\lpi}_{WW}}{R^{\lpi}_{\g\g}}  & = \frac{2}{s^4_W}\Bigg(\frac{r\sin\psi}{1 + r \cos\psi + r\sin\psi}\Bigg)^2\,.
	\end{split}
\end{equation}
(Analogous expressions for $R^{\hpi}_{VV}/R^{\hpi}_{\g\g}$ can be obtained from eqs.~\eqref{eqn:PWRR} under the replacement $r \to -[(c_1^2 + c_2^2)/c^2_{\tPi}]/r$.) The parameters $r$ and $\psi$ may therefore, for instance, be extracted from (future) measurement of $\lpi \to ZZ$ and $Z\gamma$ rates, providing an immediate prediction for $R^{\lpi}_{WW}/R^{\lpi}_{\g\g}$. Alternatively, if all four $\lpi$ decay modes are observed, a consistent global fit in the $r$--$\psi$ polar plane under eqs.~\eqref{eqn:PWRR} is a generic prediction -- a necessary condition -- of the triplet-singlet framework. 

Let us now examine the prospects for testing or excluding the singlet, triplet and triplet-singlet frameworks at the LHC, using the two-parameter relations~\eqref{eqn:PWRR}. Note that including measurements of $\hpi$ decays would potentially allow us to probe the triplet-singlet framework more deeply than measurements of $\lpi$ decays alone: In principle, including the $\hpi$ constraints allows direct measurement of $\tan\varphi$, thus lifting the projection onto the $r$--$\psi$ polar plane in eqs.~\eqref{eq:polardef}. We consider here, however, the phenomenology of only $\lpi$ decays. This permits a simpler representation of the parameter space for the branching ratio relations, and also corresponds to a `worst case' scenario, in which, for example, non-observation of $\hpi$ modes occurs because of dilution by an invisible $\hpi$ width. As such we consider the discussion and projected sensitivities presented in this section to be more conservative and model independent. 

\begin{figure}[t]\centering
\includegraphics[width=0.4\linewidth]{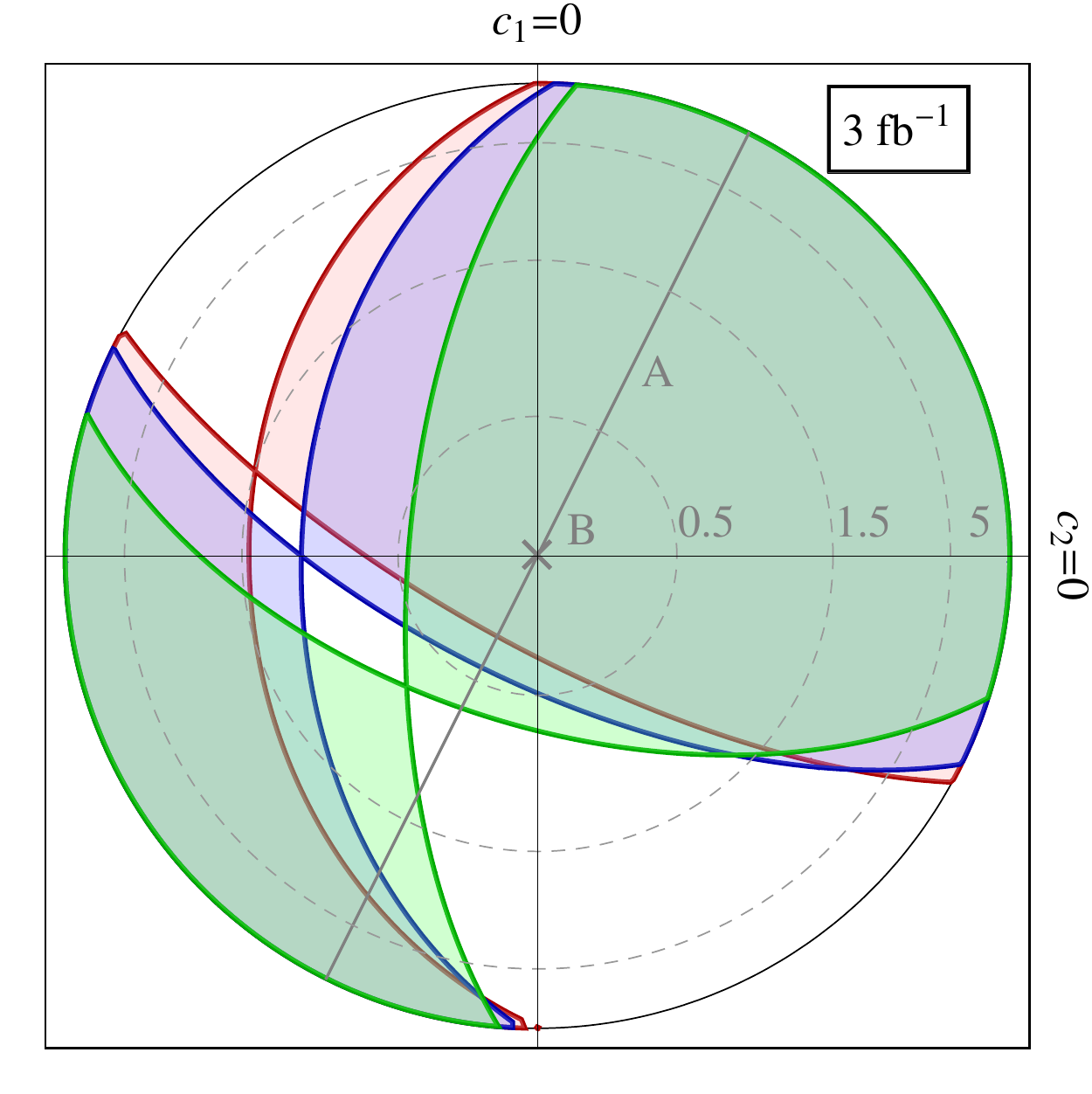}\hspace{1cm}
\includegraphics[width=0.4\linewidth]{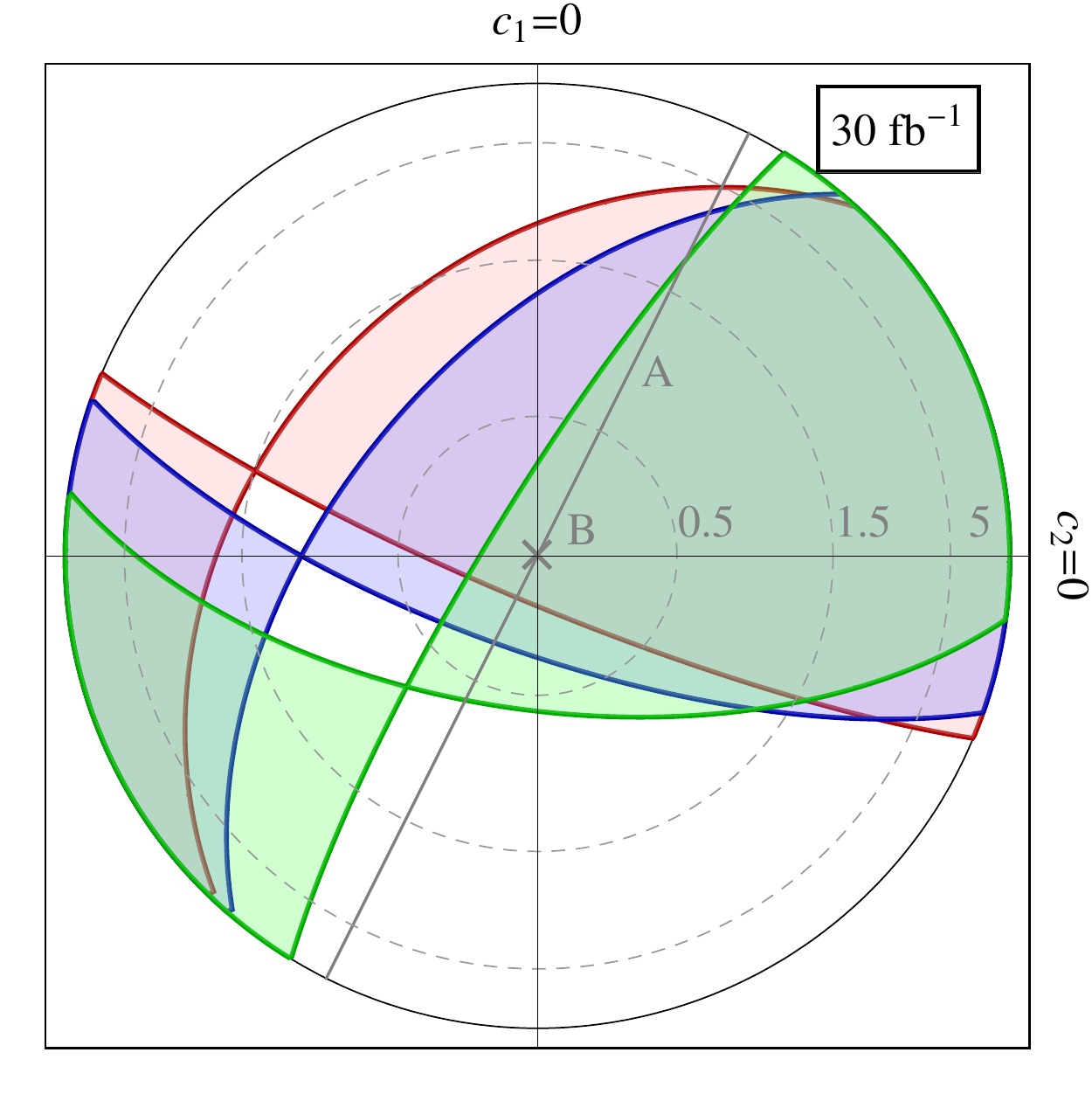} \\\vspace{1cm}
\includegraphics[width=0.4\linewidth]{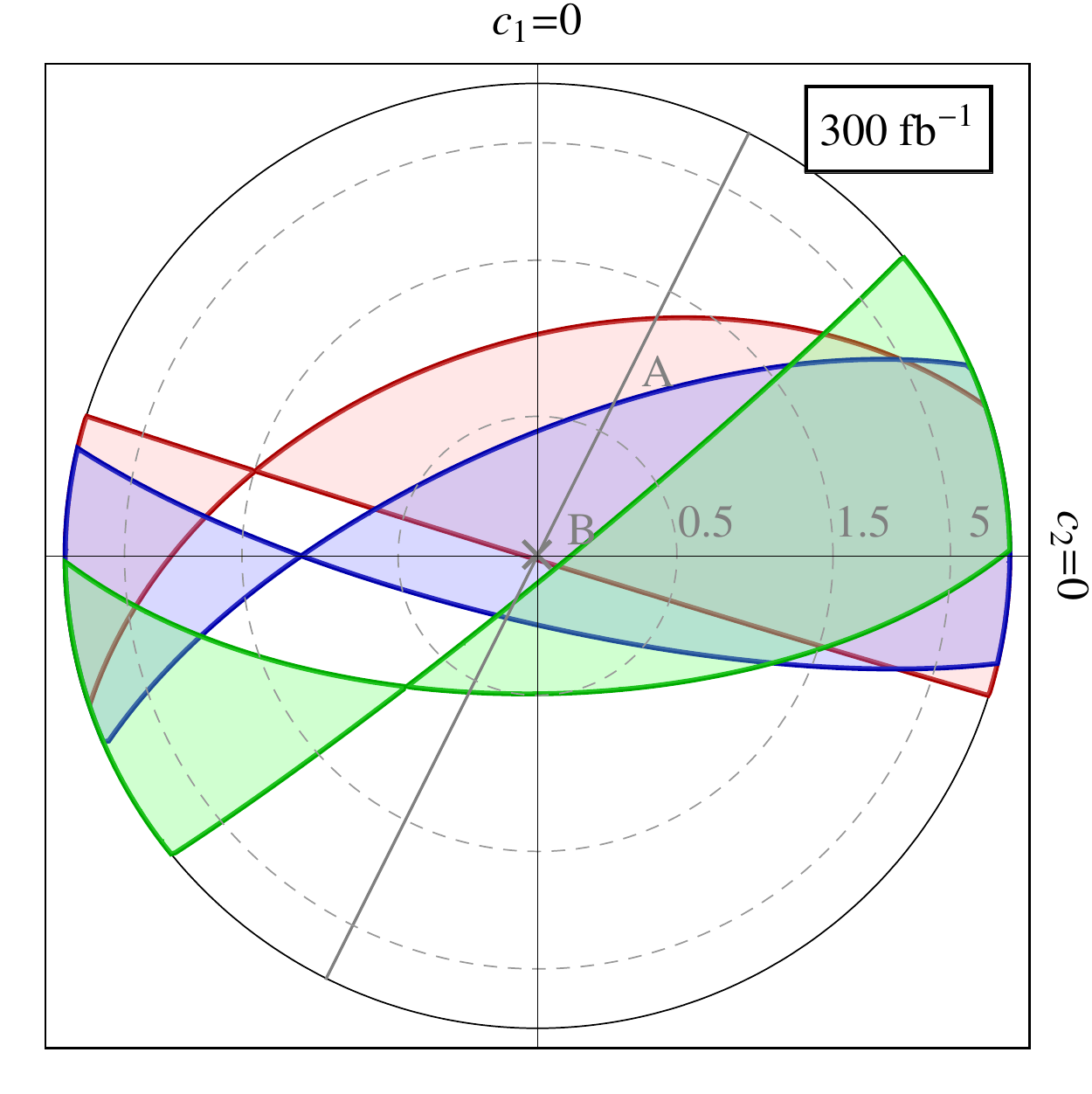}\hspace{1cm}
\includegraphics[width=0.4\linewidth]{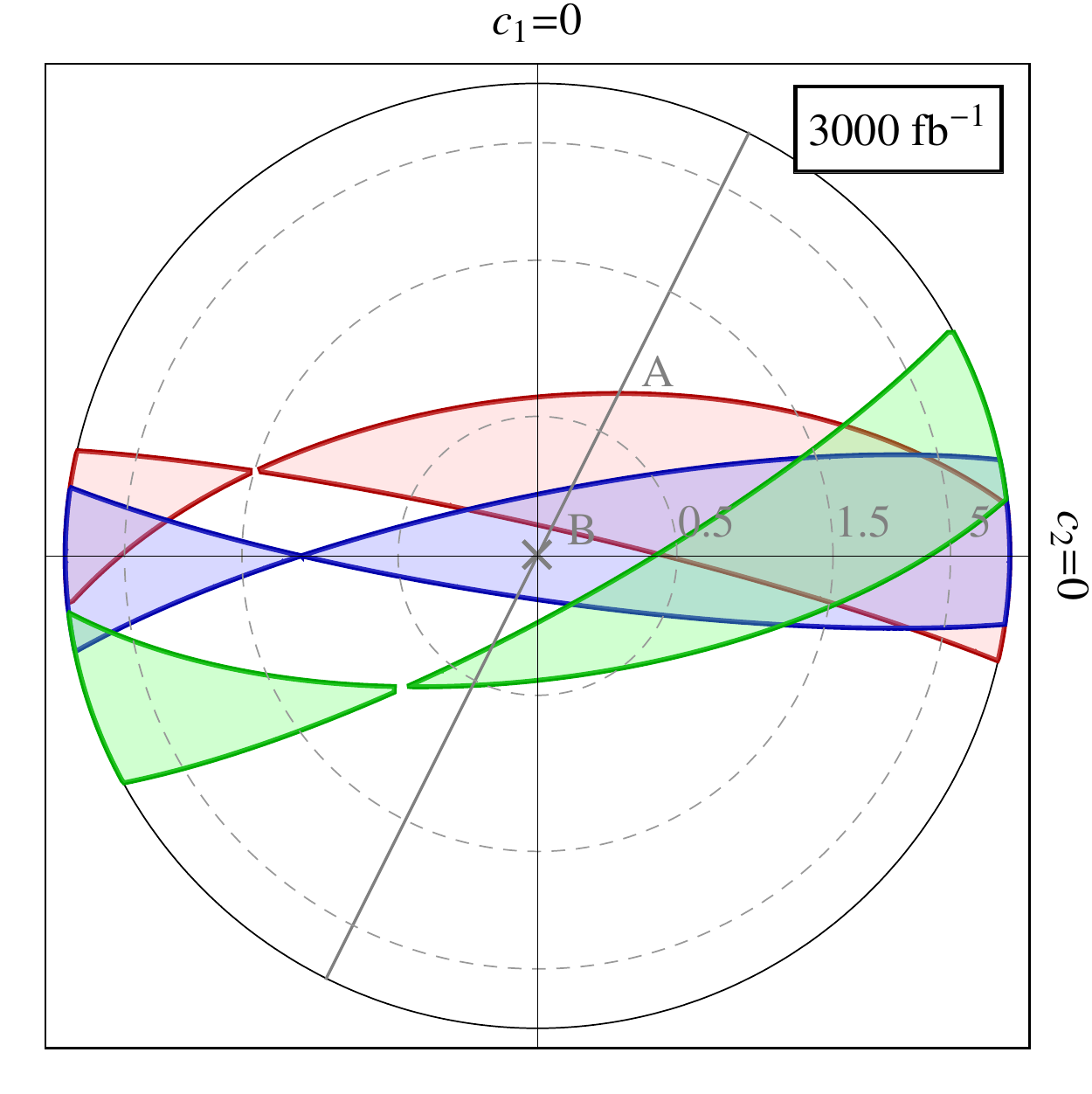}
\caption{Relative branching ratio allowed regions for resolved resonances in the $\tan^{-1}(r)$--$\psi$ polar plane for $R^{\lpi}_{VV}/R^{\lpi}_{\g\g}$, with $VV = ZZ$ (red), $Z\g$ (green) and $WW$ (blue). The origin corresponds to the pure triplet case, while the boundary of the disk corresponds to the pure singlet parameter space. Contours of constant $r$ are indicated by dashed gray lines. Also shown are the benchmark models A (grey line, parametrized by varying $\tan\varphi$) and B (grey cross).}
\label{fig:BRF}
\end{figure}

To visualize the constraints on $\lpi$, we map the infinite $r$--$\psi$ parameter space to a compact disk of radius $\pi/2$ under the conformal transformation $r \mapsto \tan^{-1}(r)$, as shown in Fig.~\ref{fig:BRF}. We emphasize that one may smoothly transition to pure singlet branching ratio relations in eqs.~\eqref{eqn:PWRR} by sending $r \to \infty$  (i.e. $\varphi \to \pi/2$, or $c_{\tPi}/\sqrt{c^2_1 + c_2^2} \to 0$). The boundary of this disk now corresponds to the one-dimensional parameter space of the pure singlet, and the ratio ${c_2}/{c_1} = \tan\psi$ then parametrizes the pure singlet branching ratios relations. The opposite limit that $r \to 0$ (i.e. $\varphi \to 0 $ or $c_{1,2} \to 0$) transitions eqs.~\eqref{eqn:PWRR} to pure triplet branching ratio relations. The origin of the disk therefore corresponds to the zero dimensional parameter space of the pure triplet.
 
In Fig.~\ref{fig:BRF} we show the allowed regions for the cases that $R^{\lpi}_{VV}/R^{\lpi}_{\g\g}$, $VV = ZZ$, $Z\g$ and $WW$, are bounded by the constraints of Table~\ref{tab:bounds}. The benchmark models A and B from Sec.~\ref{sec:diphoton} are indicated with a grey line, parametrized by varying $\tan\varphi$, and a grey cross respectively. To estimate the future sensitivity, we assume that $ZZ$, $Z\g$ and $WW$ are bounded rather than observed, and we scale the bounds in Table~\ref{tab:bounds} by $\sqrt{\mathcal{L}/\mathcal{L}_0}$ for current equivalent luminosity $\mathcal{L}_0 = 3$\,fb$^{-1}$ and future luminosities $\mathcal{L} = 30$, $300$ and $3000$\,fb$^{-1}$, corresponding to the ultimate (high luminosity) LHC reach. The net allowed region is the intersection of all three allowed regions. If there is no overlap at the origin (boundary), the pure triplet (singlet) case is excluded. The triplet-singlet scenario itself is excluded if there is no common allowed region for the three rate ratios anywhere in the $\tan^{-1}(r)$--$\psi$ parameter space.

We see in Fig.~\ref{fig:BRF} that neither the pure triplet nor pure singlet cases are excluded at $\mathcal{L}=3$\,fb$^{-1}$. Moreover, a large amount of parameter space for $r \sim \mathcal{O}(1)$ is still allowed for the triplet-singlet case. This remains the case at $30 \text{fb}^{-1}$. However, at $300$\,fb$^{-1}$ the pure triplet case (benchmark B) as well as benchmark A are excluded. At $3000$\,fb$^{-1}$ the pure singlet is just excluded, but the triplet-singlet survives in this resolved resonances scenario with $r \sim\mathcal{O}(1)$. 

\subsection{Unresolved resonances}
If the $\lpi$ and $\hpi$ resonances are unresolved, the relations~\eqref{eqn:PWRR} no longer apply, and the parameter space is in general three-dimensional. However, if we wish to fake a single, broad resonance, both $\lpi$ and $\hpi$ should have a sufficiently large coupling to gluons, in order to ensure a large enough production cross-section. (Note that this is a necessary but not sufficient condition for the interpretation of $\lhpi$ decays as a single, broad diphoton resonance, since in certain mixing angle regimes one resonance may be suppressed by destructive interference (see Fig.~\ref{fig:C1}).) It is therefore a reasonable assumption that the total widths of both resonances are dominated by the dijet mode, so that $\Br^{\lhpi}_{VV} \simeq \Gamma^{\lhpi}_{VV}/\Gamma^{\lhpi}_{gg}$. The ratios of effective rates \eqref{eqn:URR} then reduce to
\begin{equation}
	\label{eqn:MRBRR}
	\frac{R^{\text{eff}}_{VV}}{R^{\text{eff}}_{\g\g}} \simeq \frac{\Gamma^{\lpi}_{VV} + \Gamma^{\hpi}_{VV}}{\Gamma^{\lpi}_{\g\g} + \Gamma^{\hpi}_{\g\g}}\,
\end{equation}
where we also took $m_{\lpi}\simeq m_{\hpi}$. In contrast to the resolved resonance case in Sec.~\ref{sec:BRRR}, here information from both $\lpi$ and $\hpi$ decay modes are encoded into relations for $R^{\text{eff}}_{VV}/R^{\text{eff}}_{\g\g}$, by construction. The dependence on the mixing angle cancels from the ratios~\eqref{eqn:MRBRR}, such that they can again be expressed in terms of just the two polar parameters, as in eqs.~\eqref{eqn:PWRR}. Similar to eqs.~\eqref{eq:polardef}, these parameters are defined as
\begin{equation}
	 \bar{r}\cos\bar\psi \equiv c_{1}/c_{\tPi},\qquad \bar{r}\sin\bar\psi \equiv c_{2}/c_{\tPi}\,,\quad \text{so} \quad \bar r \equiv \bigg|\frac{\sqrt{c_1^2 + c_2^2}}{c_{\tPi}}\bigg|\,.
\end{equation}
The interpretation of $\bar r$ and $\bar \psi$ is the same as the interpretation of $ r$ and $ \psi$ for the resolved resonances case. The effective branching ratios of the unresolved resonances are then
\begin{equation}
	\label{eqn:PWUR}
	\begin{split}
	\frac{R^{\text{eff}}_{ZZ}}{R^{\text{eff}}_{\g\g}}  & = \frac{1 + (\bar{r}\cos\bar\psi\tan^2\theta_W + \bar{r}\sin\bar\psi\cot^2\theta_W)^2}{1 + (\bar{r}\cos\bar\psi + \bar{r}\sin\bar\psi)^2}\,,\\
	\frac{R^{\text{eff}}_{Z\g}}{R^{\text{eff}}_{\g\g}}  & = 2\frac{\cot^2(2\theta_W) + (\bar{r}\sin\bar\psi\cot\theta_W - \bar{r}\cos\bar\psi \tan\theta_W)^2}{1+ (\bar{r}\cos\bar\psi + \bar{r}\sin\bar\psi)^2}\,,\\
	\frac{R^{\text{eff}}_{WW}}{R^{\text{eff}}_{\g\g}}  & = \frac{2}{s^4_W} \frac{\bar{r}^2\sin^2\bar\psi}{1 + (\bar{r}\cos\bar\psi + \bar{r}\sin\bar\psi)^2}\,.
	\end{split}
\end{equation}

\begin{figure}[t]
\includegraphics[width=0.4\linewidth]{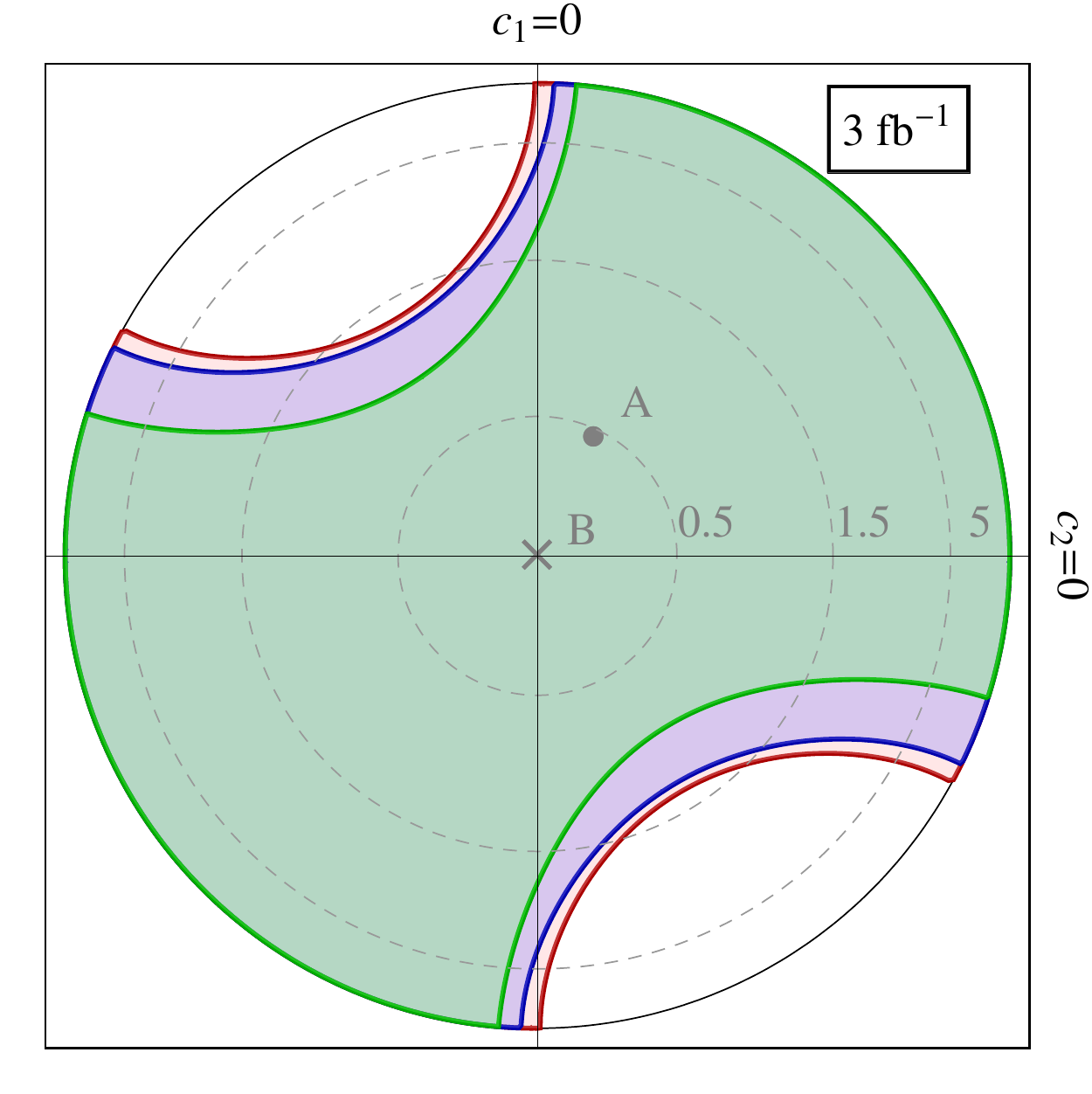}\hspace{1cm}
\includegraphics[width=0.4\linewidth]{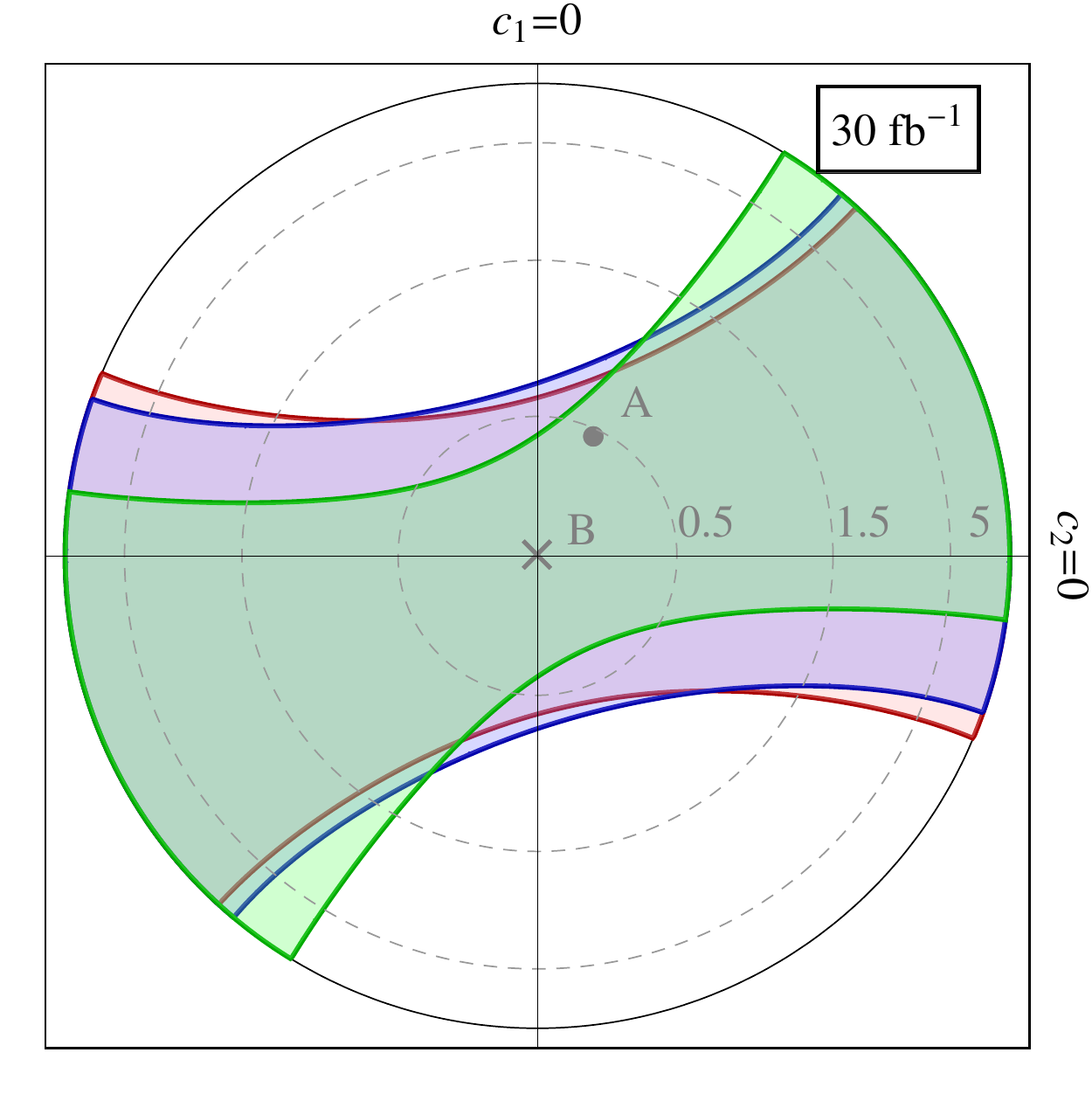} \\\vspace{1cm}
\includegraphics[width=0.4\linewidth]{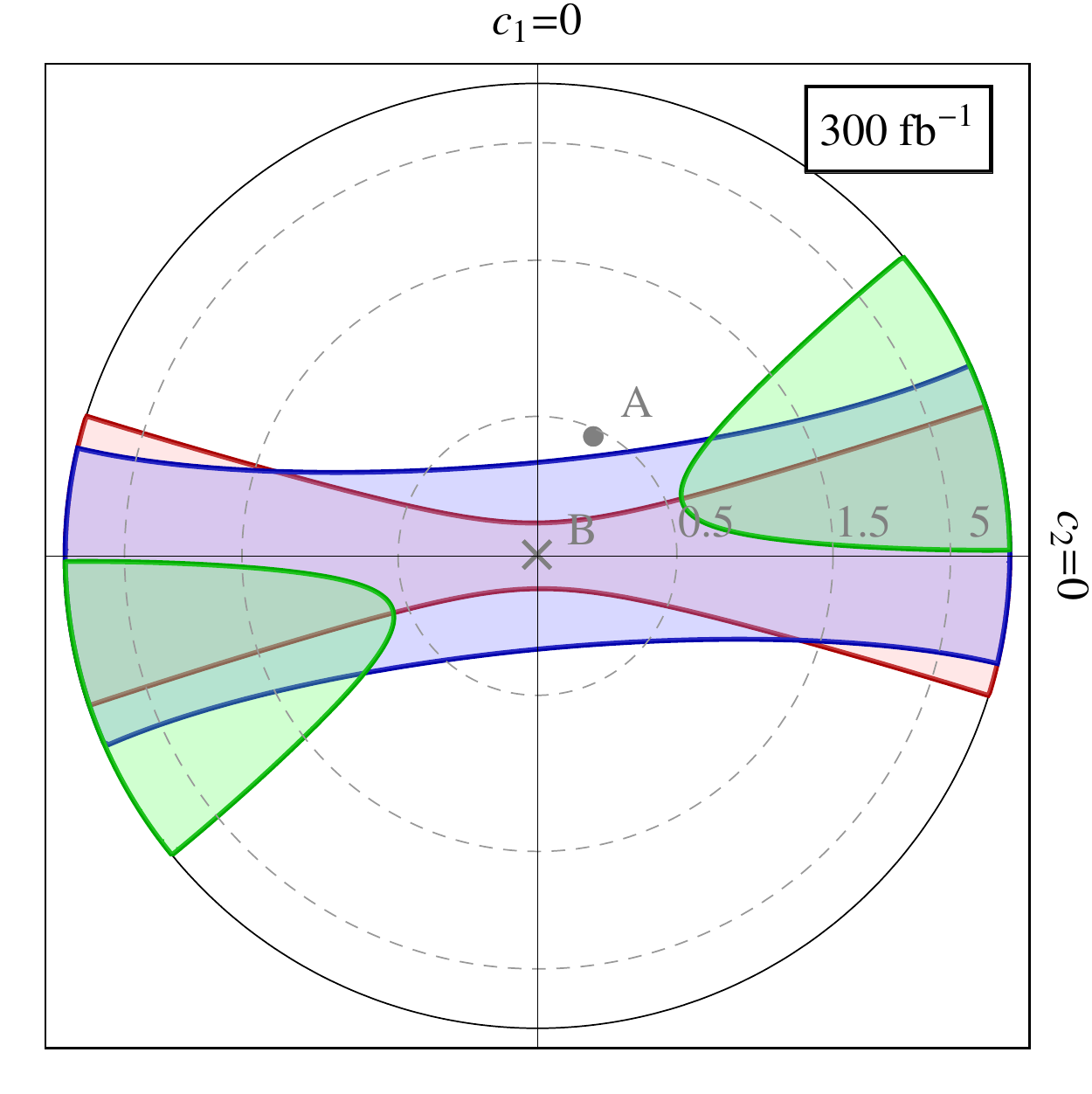}\hspace{1cm}
\includegraphics[width=0.4\linewidth]{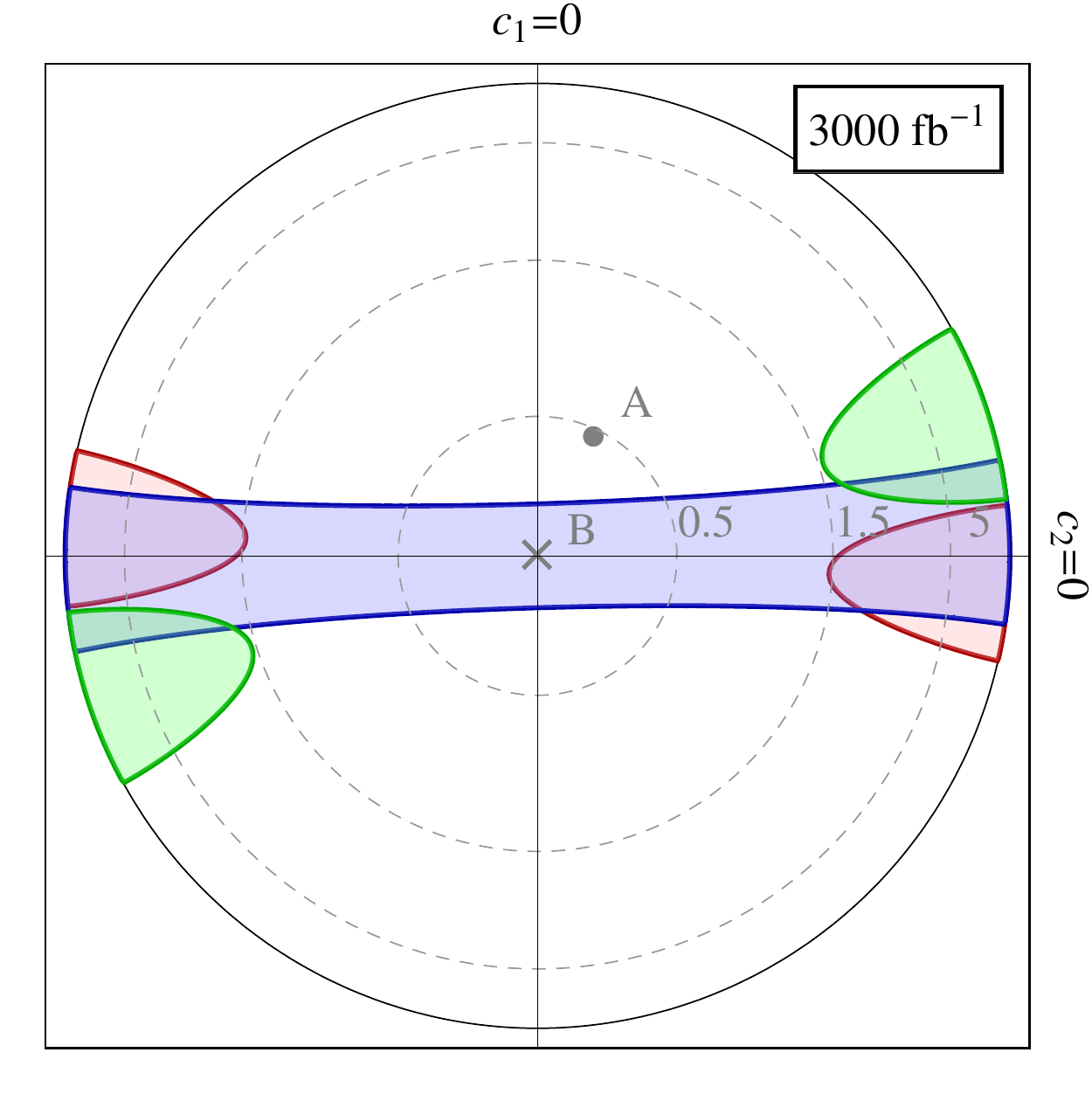}
\caption{Relative branching ratio allowed regions for unresolved resonances in the $\tan^{-1}(\bar{r})$--$\bar{\psi}$ polar plane for $R^{\text{eff}}_{VV}/R^{\text{eff}}_{\g\g}$, with $VV = ZZ$ (red), $Z\g$ (green) and $WW$ (blue). The origin corresponds to the pure triplet case, while the boundary of the disk corresponds to the pure singlet parameter space. Contours of constant $r$ are indicated by dashed gray lines. Also shown are the benchmark models A (grey dot) and B (grey cross).}
\label{fig:BRFU}
\end{figure}

Using these relations, in Fig.~\ref{fig:BRFU} we show the allowed regions for the (future) cases that the ratios $R^{\text{eff}}_{VV}/R^{\text{eff}}_{\g\g}$ are bounded by the constraints of Table~\ref{tab:bounds}, again rescaled toward $30$, $300$ and $3000$\,fb$^{-1}$. As for Fig.~\ref{fig:BRF}, we again plot under the conformal map $\bar{r} \mapsto \tan^{-1}(\bar{r})$ plane, with the understanding that the origin encodes the pure triplet point $\bar{r} \to 0$, and the $\bar{r} \to \infty$ boundary -- the boundary of the disk -- encodes the pure singlet parameter space. The benchmark models A and B from Sec.~\ref{sec:diphoton} are indicated with a grey dot and a grey cross respectively. 

With the current data and with a future luminosity of $30$\,fb$^{-1}$, neither the pure triplet, nor pure singlet, nor general triplet-singlet cases can be excluded. However, with $300$\,fb$^{-1}$ it should be possible to exclude the pure triplet -- including benchmark B -- as well as benchmark A. Notably, comparing with $30$\,fb$^{-1}$, for benchmark A we see that at least hints of all four decay modes should have been seen between $30$ and $300$\,fb$^{-1}$. At $3000$\,fb$^{-1}$, both the pure singlet and the full triplet-singlet parameter spaces are excluded, remarkably by $ZZ$ and $Z\g$ alone.  By comparison to Fig.~\ref{fig:BRF}, however, the resolved resonances triplet-singlet case is not disfavored at this luminosity. We thus see that one may, in principle, use the relations~\eqref{eqn:PWRR} and \eqref{eqn:PWUR} to disfavor the unresolved versus resolved resonances cases, even if the three other neutral diboson modes are not observed at LHC.

\section{Composite models}
\label{sec:composite}

Composite models motivate the presence of new light scalars, without introducing a new hierarchy problem. The pseudo-Nambu-Goldstone bosons (pNGBs) of dynamically broken chiral symmetries are particularly attractive candidates for the EW triplet and singlet pseudoscalars, $\tPi$ and $\szero$, as they can naturally be separated from the scales of other new composite states, and a wide class of composite sectors contain both triplet and singlet pNGBs. 

In this class of theories, charging hyperquarks (hereafter we identify states in the composite sector by their SM analogs, adding a `hyper' prefix) under SM gauge groups, or more generally embedding SM gauge groups into the global symmetries of the composite sector, is the leading portal to the SM sector~\cite{Kilic:2008pm,Kilic:2009mi,Bai:2010mn}. Dimension-five couplings of the pNGBs to SM gauge bosons are then generated by chiral anomalies. In the absence of mixing between the SM and composite fermions, the Higgs portal is the next leading coupling into the composite sector. In particular, the quartic operators~\eqref{eq:quartics}, that mix an EW triplet and singlet pNGB, can be generated when the Higgs couples to the composite sector. 

In the context of an $SU(N_c)$-type confining theory, the effective field theory scale $f$ encodes the chiral symmetry breaking scale $\Lambda \sim 4\pi f/\sqrt{N_c}$, that corresponds to the mass scale of the other composite states in the spectrum. The chiral anomaly coefficients for such a theory determine the viable range of $f$ to produce the observed signatures, as in Section~\ref{sec:diphoton} above. The most relevant heavy states for collider phenomenology are the hyper-$\rho$ vector mesons, that mix with the SM gauge bosons and dominantly decay to the hyper-pions~\cite{Kilic:2008pm,Kilic:2009mi,Kilic:2010et}. We focus on on the case that $f\gtrsim 500$ GeV and $N_c \sim 5$, so that $m_{\rho} \sim 4\pi f/\sqrt{N_c} \gtrsim 2.5$ TeV. In this case, the hyper-$\rho$'s lead to at most an $\mathcal{O}(1)$ enhancement over the Drell-Yan $\tpm\tzero$ or $\tpm\tmp$ pair production rates. However, the single-production rate of $\lhpi$ remains much larger than those of the pair production processes.

We first describe the model of Ref.~\cite{Harigaya:2016pnu}, which presents a simple hidden sector QCD-like theory containing both a singlet and triplet pNGB, without any couplings to the Higgs. This is a useful benchmark, as Ref.~\cite{Harigaya:2016pnu} has shown that the singlet can reproduce the diphoton anomaly and that the extra exotic states in the model can be made cosmologically safe and consistent with collider observations. We introduce here couplings of the composite sector to the Higgs, mixing the triplet and singlet and thereby significantly modifying the phenomenology of the theory. We then discuss some simple variations, the general conditions for a triplet to emerge from the compositeness sector, and some interesting new features in models where not only the triplet and singlet, but also Higgs itself emerges as a pNGB from the composite sector.

\subsection{Benchmark composite model}
\label{sec:benchmark}

The model of Ref.~\cite{Harigaya:2016pnu} is a QCD-like $SU(N_c)$ hypercolor gauge theory, with an $SU(5)_F$ flavor group. Vector-like hyperquarks $\Psi$ and $\bar\Psi$ transform under the $(\Box\,,\,\bm{5})$ and $(\bar\Box\,,\,\bar{\bm{5}})$ of $SU(N_c)\times SU(5)_F$. In the massless limit, the hyperquarks exhibit accidental $SU(5) \times SU(5)$ global symmetries that break to the diagonal $SU(5)_F$ under $SU(N_c)$ confinement, producing one singlet heavy hyper-$\eta'$ meson, with mass $\sim \Lambda$, and 24 pNGBs in the adjoint of $SU(5)_F$.

Embedding the gauged SM into $SU(5)_F$, such that $SU(5)_F\supset SU(3)_C\times SU(2)_L\times U(1)_Y$, it is convenient to decompose $\Psi$ into SM irreps, writing $\Psi=(\Psi_d, \Psi_\ell)$, with $\Psi_d \sim (\Box, \bar{\bm{3}}, \bm{1},1/3)$ and $\Psi_\ell \sim (\Box, \bm{1}, \bm{2},-1/2)$ under $SU(N_c) \times$\,SM, and similarly for $\bar\Psi$. Under the decomposition of the pNGB adjoint, the condensate $\Psi_\ell \bar \Psi_\ell$ contains the triplet pseudoscalar pNGB, $\tPi \sim (\bm{1},\bm{3},0)$, while the singlet $\szero \sim (\bm{1},\bm{1},0)$ comes from the non-anomalous singlet combination in $\Psi_\ell \bar \Psi_\ell$ and $\Psi_d \bar \Psi_d$, i.e. $\szero \sim \sqrt{3/5}\big(\Psi_d\bar\Psi_d/3 - \Psi_\ell \bar\Psi_\ell /2 \big)$. This theory also contains colored pNGBs, a complex $SU(3)_c$ triplet $\chi \sim (\bar{\bm{3}},\bm{2},5/6)$ and an octet $\psi \sim (\bm{8},\bm{1},0)$. The phenomenology of these colored states has been discussed in Ref.~\cite{Harigaya:2016pnu}, and will be unchanged by the Higgs portal couplings that we will introduce (see also Ref.~\cite{Cacciapaglia:2015eqa} for colored pNGBs in composite Higgs models).

The low energy theory is as described in \Sref{sec:framework}, with $f$ the decay constant scale and the anomaly coefficients fixed as
\begin{gather}
	\label{eqn:ACBT}
	c_{\hat\Pi} = N_c\,, \\
	c_1 = \sqrt{\frac{3}{5}}\frac{5}{18} \simeq 0.2 N_c\,,\qquad c_2 = \sqrt{\frac{3}{5}}\frac{1}{2} N_c \simeq 0.4 N_c\,,\qquad c_3 = -\sqrt{\frac{3}{5}}\frac{1}{3} N_c \simeq -0.25 N_c\,.\notag
\end{gather}
As in Ref.~\cite{Harigaya:2016pnu}, the pNGB masses are generated by both the gauging of the SM gauge group and explicit mass terms for the hyper-quarks, 
\begin{equation}
	M_\ell \Psi_\ell \bar \Psi_\ell + M_d \Psi_d \bar \Psi_d\,.
\end{equation}
This gives na\"\i ve dimensional analysis (NDA) estimates for the pNGB masses~\eqref{eqn:MTE}
\begin{equation}
	\label{eqn:MNDA}
	m^2_\tPi \simeq 2 M_\ell \Lambda + \frac{6 g_2^2 f^2}{N_c}\,, \qquad m^2_\szero \simeq \frac{(6 M_\ell + 4 M_d)}{5}\Lambda\,.
\end{equation}
From eqs.~\eqref{eqn:C1DR} or \eqref{eqn:C2DR}, the diphoton rate can be fit for $f\sim\TeV$ and $N_c\simeq5$. For triplet and singlet masses both near $750~\GeV$, the NDA estimates~\eqref{eqn:MNDA} suggest that the triplet mass can be primarily generated by the gauge contributions, while we are free to set $M_\ell\sim 0$. In this limit, the colored partners have masses at $m_\chi \gtrsim 1.0$ TeV and $m_\psi \gtrsim 1.5$ TeV, beyond current LHC bounds \cite{Harigaya:2016pnu}. For larger scales $f\gtrsim \TeV$, the radiative corrections $\sim g_2^2$ must be smaller than the expected NDA size to obtain $m_\tPi\sim 750~\GeV$.

Besides the gauging of (subgroups of) $SU(5)_F$, the lowest dimension portal between the SM and hypercolor sectors is through the dimension-five Higgs portal operators (cf. eqs.~\eqref{eq:quartics})
\begin{equation}
	\label{eqn:HP}
	\frac{1}{\LambdaP}\bigg[\hat \lambda \big(H^\dagger \sigma^a H\big)\big(\Psi_\ell \sigma_a \bar \Psi_\ell\big) + \hat \lambda_\ell |H|^2 \Psi_\ell \bar \Psi_\ell +  \hat \lambda_d |H|^2 \Psi_d \bar \Psi_d\bigg]\,.
\end{equation}
The coupling $\hat \lambda$ breaks the custodial symmetry and will generate the $\tzero$--$\szero$ mixing. We can make an NDA estimate of the size of these operators compared to the effective operators~\eqref{eq:quartics} in the low energy description, yielding $\lambda \sim \hat\lambda\Lambda/\LambdaP$, $\lambda_\tPi\sim \hat\lambda_\ell \Lambda/\LambdaP$, $\lambda_\szero \sim \hat\lambda_{\ell,d} \Lambda/\LambdaP$.  If the hypercolor sector is asymptotically free, these operators are irrelevant in the UV theory, suggesting a UV completion at scale $\LambdaP$. Perturbativity at $\LambdaP$ then requires $\LambdaP/\Lambda \lesssim 16\pi^2/\lambda$. The simplest such completion involves an extra singlet hyperquark at $\Psi_S$ with mass $M_S\sim\LambdaP$ generating a Yukawa portal $H^\dagger\Psi_\ell \bar\Psi_S + \Psi_S \bar\Psi_\ell H$. (For $M_S \lesssim \Lambda$, this can be viewed as a theory where the Higgs mixes with a composite doublet.) Note that the portal operators break the chiral symmetries and give natural scales for the hyperquark masses,
\begin{equation}
	\Delta M_\ell \sim \frac{\LambdaP}{16\pi^2}\bigg(\hat\lambda_{\ell} + \frac{g_2^2\hat\lambda}{16\pi^2}\bigg)\,,\qquad \Delta M_d \sim \frac{\hat\lambda_{d} \LambdaP}{16\pi^2}\,.
\end{equation}
It follows that the triplet and singlet pNGB masses correspondingly acquire mass contributions
\begin{equation}
\label{eqn:UVtriplet}
	\Delta m^2_{\tPi} \sim \frac{\LambdaP^2}{16 \pi^2}\bigg(\lambda_{\tPi} +  \frac{g_2^2\lambda}{16\pi^2}\bigg)\,, \qquad \Delta m^2_{\szero} \sim \lambda_{\szero} \frac{\LambdaP^2}{16 \pi^2}\,.
\end{equation}
Requiring that \eqref{eqn:UVtriplet} gives a contribution to the triplet mass smaller than the NDA IR gauge contribution of~\eqref{eqn:MNDA}, we find $\LambdaP/\Lambda \lesssim 4\pi/\sqrt{\lambda}$, which is more stringent than the perturbativity constraint at $\LambdaP$. We also require $\lambda_{\tPi,\szero}\ll1$, but these couplings were already not relevant for the phenomenology we have studied above.

An alternative to an asymptotically free hypercolor theory is a  theory that remains near a strongly interacting fixed point above $\Lambda$ with large anomalous dimensions for the fermion bilinears. The scale $\LambdaP$ can be pushed arbitrarily high as the scaling dimension of $\Psi\bar\Psi$ goes to $2$, although a mechanism is still needed to cut off the contributions $\Delta M_{d,\ell}$ in this case (Ref.~\cite{Harigaya:2016pnu} discusses some of the other advantages of such a UV completion). 

Apart from the Higgs portal~\eqref{eqn:HP}, there are also generically higher dimension interactions with the SM, in particular the dimension-7 Yukawa portals of the form
\begin{equation}
	\frac{\lambda_u}{\LambdaP^3} \big(H^\dagger \sigma_a \bar{Q}_L\big) u_R \big(\Psi_\ell \sigma_a \bar \Psi_\ell\big)\,,
\end{equation}
where $\lambda_u$ is the usual SM Yukawa coupling: We assume the presence of a minimal flavor violation or flavor alignment mechanism, to avoid dangerous flavor violating effects. These operators may generate the dimension-5 operators $\lambda_u H^\dagger \tPi \bar{Q}_L u_R$ or $\lambda_u \szero H \bar{Q}_L u_R$ and so on, in the low energy effective theory. Such operators are, however, heavily suppressed by $1/\LambdaP^3$ and NDA factors, producing negligible partial widths for $\lpi \to t\bar{t}$ or other fermionic decay modes, compared to the diboson partial widths generated by the chiral anomalies.

In addition to the renormalizable effective operators involving the pNGB fields, there will be contributions to the T-parameter from the heavy composite states at the scale $\Lambda$. NDA estimates for their size give $c_{T,UV}\sim\lambda^2 N_c/\Lambda^2$, which is typically subdominant to the IR contribution calculable in the effective theory \eqref{eq:Tparameter}: $c_{T,UV}/c_{T,IR} \sim N_c m^2_{\szero}/\Lambda^2$. If CP is not conserved in the hidden sector, a direct tree-level Higgs-triplet mixing operator is also generated $\sim \lambda \theta_{\rm CP} f (H^\dagger \tPi H)$. This is dangerous for electroweak precision, and requires $\theta_{\rm CP} \lesssim 1/(4\pi)$ to be subdominant to the effects of the loop-level $c_{T,IR}$. This can be natural if $\theta_{CP}$ is small because of a UV symmetry \cite{Draper:2016fsr} or if there is an axion in the hypercolor sector. This occurs, for example, when the Yukawa portal singlet obtains its mass dynamically via $\langle S \rangle \Psi_S \bar \Psi_S$. Setting $M_d \ll M_\ell$ or introducing an $SU(2)$ singlet with $M_S \ll M_\ell$ to suppress the effects of $\theta_{CP}$ is also possible, but leads to additional light mesons unless there are large four-fermion operators lifting their masses. 

\subsection{Other composite models}

Moving beyond the benchmark model, there are a wide variety of possibilities for composite sectors that reduce to the effective triplet-singlet theory. Any vector-like theory containing $SU(2)$-charged hyperquarks in a complex representation of the hypercolor group will contain triplets pNGBs, as will any hyperquarks in a real representation with a symmetric bilinear. (An example of a theory containing hyper-quarks charged under SU(2) and no triplets is the $SU(4)/Sp(4)$ coset obtained from an anti-symmetric condensate of an Sp(N) gauge group \cite{Barnard:2013zea}.) When colored hyperquarks are also present, there is also generically a singlet pNGB with an $SU(3)_c$ anomaly. (Ref.~\cite{Bai:2010mn} is an example of a theory in which an $SU(2)$ triplet is generated without any light singlets, although simple extensions of this model contain singlets along with additional gauge charged mesons.) 

Depending on the choices of gauge groups and representations, the anomaly structure can be modified from the benchmark theory, although constraints on new stable or long-lived states constrain the possibilities. A particular interesting possibility is to embed the hyperquarks in higher representations of the hypercolor $SU(N_c)$ gauge group, with dimension $d_R$. Then $N_c\rightarrow d_R$ in the anomaly coefficients~\eqref{eqn:ACBT}. For example, for $N_c=5$ the symmetric two-index irrep has $d_R=15$, which increases rates by a factor of $(15/5)^2=9$. When $\Psi_d$ and $\Psi_\ell$ are in different irreps, some of the colored mesons are removed and new colored fermions may exist, just as in the composite Higgs models of Refs.~\cite{Barnard:2013zea, Cai:2015bss}. 

Two further interesting and qualitatively different possibilities are to embed the triplet in a sector with an approximate custodial symmetry broken only softly, and to embed the Higgs itself in the composite sector.

\subsubsection{Softly broken custodial symmetry}

In larger cosets, triplets can emerge and couple to the Higgs in a custodial symmetry preserving manner. A soft explicit breaking of the custodial symmetry can then generate the singlet-triplet mixing. For example, considering only the electroweak sector, the coset $SO(4)\times SO(4)' \rightarrow SO(4)_C \simeq SU(2)_L\times SU(2)_R$ gives a pNGB sector containing both a triplet $\pi_{L}$ of $SU(2)_L$ and a triplet $\pi_R$ of the unbroken custodial $SU(2)_R$. This global symmetry structure can in principle be realized in a QCD-like hyperquark theory with a $SU(4)\times SU(4)'$ global symmetry explicitly broken to $SO(4)\times SO(4)'$ by four-fermion operators.

The Higgs can couple to this sector in an $SO(4)_C$-preserving fashion, giving a mixing between $\pi_L$ and $\pi_R$. As $U(1)_Y$ breaks $SU(2)_R$, anomalies can generate a coupling $\pi_R^0 B_{\mu\nu}\tilde{B}^{\mu\nu}$ without any other sources of explicit $SU(2)_R$ breaking in the composite sector (this is analogous to the $\pi^3_L W^3_{\mu\nu}\tilde B^{\mu\nu}$ coupling). However, a large coupling $\pi^0_R G_{\mu\nu}\tilde{G^{\mu\nu}}$ requires further explicit breaking of $SU(2)_R$, which can be achieved by operators mixing a singlet pNGB $\eta^*$ with a $\pi^0_R$. (Another possibility is that colored and uncolored hyperquarks combine into $SU(2)_R$ multiplets, so that the gauging of $SU(3)_c$ itself explicitly breaks $SU(2)_R$). In fact, in the limit that a heavier combination of $\eta^*$ and $\pi_R^0$ can be integrated out, the singlet sector of such a model reduces to the same effective theory we have described in \Sref{sec:framework}, although the charged sector may be more complicated. The IR contributions to the T-parameter scale just as for the simpler triplet-singlet model, but the UV contributions are suppressed because of the soft nature of the custodial breaking.

\subsubsection{Non-minimal composite Higgs}

If the custodial symmetry is approximately preserved by the composite sector, it is possible that the Higgs itself emerges from the composite sector as well. For example, the coset $SO(5)\times SO(5)'\rightarrow SO(5)$ contains the Higgs in a $\bm{4}$ of $SO(4)_C$ in addition to $\pi_L$, $\pi_R$. When such a model contains fermionic top partners, as required for the partial compositeness mechanism that generates the large top Yukawa coupling, there are naturally singlet pNGBs with $G\tilde{G}$ anomalies \cite{Cai:2015bss}. The triplet can mix with these states as described in the $SO(4)\times SO(4)'$ model above.

An interesting alternative possibility for generating a coupling to gluons is that the triplets couple to the top quark axial currents,
\begin{equation}
 \frac{i\partial_\mu \pi_L^a}{f} q^\dagger_3 \sigma^a \sigma^\mu q_3 \rightarrow \frac{m_t}{f} (\tzero t t^c + \tminus t b^c + \tplus t^c b) + \hc,
\end{equation}
so that gluon couplings are generated by top quark loops. The large branching ratio of $\pi^0_L$ into top quarks makes it difficult for this state to be the diphoton resonance itself, but when the triplet also has large widths for tree-level cascade decays to a lighter 750\,GeV state, this can lead to interesting phenomenology, as discussed in Sec.~\ref{sec:DCD}.

\section{Conclusions}
\label{sec:conclusions}

If the recent hints of a $750~\GeV$ diphoton resonance observed at the LHC are really the first signs of new physics, a detailed exploration will be possible with the full LHC luminosity. The simplest phenomenological possibility, a singlet scalar resonance, has a rather constrained set of observables \cite{Harigaya:2015ezk,Mambrini:2015wyu,Backovic:2015fnp,Angelescu:2015uiz,Nakai:2015ptz,Knapen:2015dap,Buttazzo:2015txu,Pilaftsis:2015ycr,Franceschini:2015kwy,McDermott:2015sck,Ellis:2015oso,Low:2015qep,Bellazzini:2015nxw,Gupta:2015zzs,Molinaro:2015cwg,Bian:2015kjt,Agrawal:2015dbf,Falkowski:2015swt,Aloni:2015mxa,Bai:2015nbs,Altmannshofer:2015xfo}, and it is therefore well-motivated to consider whether the resonance could arise from the neutral components of higher $SU(2)_L$ representations. In this work we have studied the possibility that the new physics involves a pseudoscalar electroweak triplet, that mixes with a EW singlet after EWSB. Compared to a doublet, an EW triplet may decay to diboson final states already at the dimension-five level, without requiring additional sources of EWSB; mixing with the scalar opens up a gluon fusion production channel, leading to a much richer phenomenology near 750 GeV. Apart from the diphoton resonance itself, this includes altered diboson branching ratios, cascade decays, and Drell-Yan pair production of the charged states.

The triplet-singlet mixing framework can be viewed as a concrete completion of a theory containing a single $750~\GeV$ scalar with $B_{\mu\nu}\tilde{W}_a^{\mu\nu}$, $B_{\mu\nu}\tilde{B}^{\mu\nu}$, $W^a_{\mu\nu}\tilde{W}_a^{\mu\nu}$, and $G^a_{\mu\nu}\tilde{G}_a^{\mu\nu}$ couplings ($c_\tPi$, $c_1$, $c_2$, and $c_3$). Our study of the diboson branching ratios in \Sref{sec:BRR} applies more generally to any such scenario where only these couplings are generated (although it does not apply if, e.g., the $W_{\mu\nu}^3  {\tilde W^{\mu\nu}}^3$ coupling is linearly independent from the $W_{\mu\nu}^+  {\tilde W^{\mu\nu}}^-$ coupling as can occur in some models).
The singlet-triplet model is a particularly attractive option because, as we have shown, electroweak precision observables and Higgs properties can be consistent even when the 750 GeV state has large couplings both to the triplet operator $B_{\mu\nu}\tilde{W}_a^{\mu\nu}$ and the singlet $B_{\mu\nu}\tilde{B}^{\mu\nu}$ and $W^a_{\mu\nu}\tilde{W}_a^{\mu\nu}$ operators. Composite models, in which the triplet and singlet emerge as pNGBs from a new hypercolor gauge sector, are a natural UV completion for the triplet-singlet model. In particular, the dimension-5 couplings to the SM gauge bosons are generically generated by the chiral anomalies, and the mixing of the triplet and singlet pNGB arises from the Higgs portal coupling to the composite sector. The singlet and triplet pNGB can be the lightest states and most relevant for collider phenomenology, although these models may also predict heavier colored pNGBs and hyper-$\rho$ vector mesons. We have studied in detail the phenomenology of one simple benchmark model for the composite sector, but a wide variety of possibilities exist. 

In this general triplet-singlet framework, we find that a narrow diphoton signature may be generated from the lighter of the mixed triplet-singlet neutral states, $\lpi$.  Observations consistent with a broader resonance can arise if
two neutral mass eigenstates in the triplet-singlet admixture, $\lhpi$, have a small mass splitting, such that they produce unresolved, overlapping resonances. In either case, the couplings to the SM gauge bosons cover a more general space of branching ratios to the diboson final states $\gamma\gamma$, $Z\gamma$, $ZZ$, $WW$ than is possible for a pure singlet or pure triplet.  These relations may be conveniently parametrized on a compact two-dimensional space together with the current and projected LHC reach. In both cases, with $300$\,fb$^{-1}$, LHC measurements in the other $750$\,GeV diboson channels can rule out the possibility of a pure triplet, and the pure singlet is excluded at $3000$\,fb$^{-1}$. At the latter luminosity, for narrow, resolved resonances, the mixed triplet-singlet scenario can still be consistent with non-observation of other diboson decay modes. However, in the unresolved resonances case, it is excluded by projected $ZZ$ and $Z\g$ bounds alone.

The heavier neutral state, $\hpi$, may exhibit large branching fractions to the lighter charged and neutral scalars, $\hpi \to \tmp W^\pm$, $h \lpi$. This can be the dominant production mechanism for the charged states, or even the 750 GeV state itself. Beyond the present diphoton excess motivation, such tree-level cascade decays are a generally interesting phenomenon to consider: The triplet-singlet model in particular motivates multi-resonant searches in the unusual tri-boson channels $h V V$ and $W W \gamma$ if the dominant production mechanism is gluon fusion for the heavier singlet, and the six-boson $W W (V V) (V V)$ and $W h (V V) (V V)$ channels if the cascade is initiated from Drell-Yan pair production. Finally, the triplet-singlet framework also features irreducible Drell-Yan pair production cross-sections for $\lhpi \tpm$ and $\tpm\tmp$, that decay to double diboson resonances, in the latter case with rates determined by SM electroweak couplings alone. Such signatures can probe the presence of an electroweak triplet, with the $W\gamma W\gamma$ and $W\gamma\gamma\gamma$ double diboson resonance being the most promising channels.

\textbf{Acknowledgments:} We thank Bogdan Dobrescu, Roni Harnik and Paddy Fox for the discussions which initiated this work. We are also grateful to Keiseke Harigaya, Jack Kearney,  Zhen Liu, Tim Lou, Michele Papucci, and Diego Redigolo for useful discussions. We further thank Bogdan Dobrescu, Can Kilic and Diego Redigolo for valuable comments on the manuscript. The work of SK was supported by the LDRD Program of LBNL under U.S. Department of Energy Contract No. DE-AC02-05CH11231. Fermilab is operated by Fermi Research Alliance, LLC under Contract No. \protect{DE-AC02-07CH11359} with the United States Department of Energy. The work of DR is supported by the National Science Foundation (NSF) under grant No. PHY-1002399.


%

\end{document}